\documentclass[lettersize,journal]{IEEEtran}
\usepackage{amsmath,amsfonts}
\usepackage{algorithmic}
\usepackage{algorithm}
\usepackage{array}
\usepackage[caption=false,font=normalsize,labelfont=sf,textfont=sf]{subfig}
\usepackage{textcomp}
\usepackage{stfloats}
\usepackage{url}
\usepackage{verbatim}
\usepackage{graphicx}
\usepackage{cite}

\usepackage{sparklines}
\usepackage{spark}
\usepackage{color, colortbl}
\usepackage{xcolor}
\usepackage{soul}
\usepackage{enumitem}
\usepackage{booktabs}
\usepackage{makecell}
\usepackage{makecell}

\usepackage{etoolbox}
\makeatletter
\patchcmd{\@makecaption}
  {\scshape}
  {}
  {}
  {}
\makeatother

\usepackage{hyperref}

\begin{document}

\title{InsigHTable: Insight-driven Hierarchical Table Visualization with Reinforcement Learning}

\author{Guozheng~Li, Peng~He, Xinyu~Wang, Runfei~Li, Chi~Harold~Liu, Chuangxin Ou, Dong He, and Guoren~Wang
\thanks{Guozheng~Li, Peng He, Xinyu Wang, Runfei Li, Chi Harold Liu, and Guoren Wang are with the School of Computer Science and Technology, Beijing Institute of Technology. E-mail: \{guozheng.li, hepeng, wang.xinyu, lirunfei, chiliu, wanggr\}@bit.edu.cn. Chi Harold Liu is the corresponding author. 
}%
\thanks{Chuangxin Ou and Dong He are with PICC Information Technology Company Limited.}
}

\maketitle

\begin{abstract}
Embedding visual representations within original hierarchical tables can mitigate additional cognitive load stemming from the division of users’ attention. 
The created hierarchical table visualizations can help users understand and explore complex data with multi-level attributes.
However, because of many options available for transforming hierarchical tables and selecting subsets for embedding, the design space of hierarchical table visualizations becomes vast, and the construction process turns out to be tedious, hindering users from constructing hierarchical table visualizations with many data insights efficiently.
We propose InsigHTable, a mixed-initiative and insight-driven hierarchical table transformation and visualization system.
We first define data insights within hierarchical tables, which consider the hierarchical structure in the table headers.
Since hierarchical table visualization construction is a sequential decision-making process, InsigHTable integrates a deep reinforcement learning framework incorporating an auxiliary rewards mechanism. 
This mechanism addresses the challenge of sparse rewards in constructing hierarchical table visualizations. 
Within the deep reinforcement learning framework, the agent continuously optimizes its decision-making process to create hierarchical table visualizations to uncover more insights by collaborating with analysts. 
We demonstrate the usability and effectiveness of InsigHTable through two case studies and sets of experiments. 
The results validate the effectiveness of the deep reinforcement learning framework and show that InsigHTable can facilitate users to construct hierarchical table visualizations and understand underlying data insights.
\end{abstract}

\begin{IEEEkeywords}
Hierarchical tabular data, table visualization, reinforcement learning, data transformation.
\end{IEEEkeywords}

\section{Introduction}
\label{sec:introduction}
\begin{figure}[!ht]
\centering  
\hfill
\subfloat[]{
    \includegraphics[width=0.47\columnwidth]{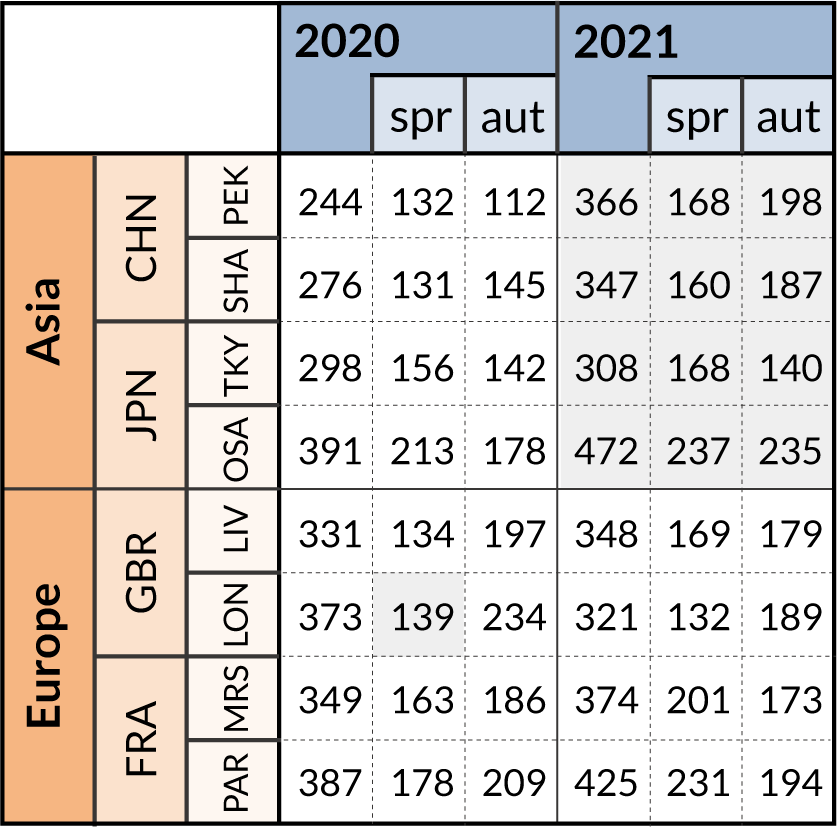}
    \label{fig:tablevis-example-a}
}\hfil
\subfloat[]{
    \includegraphics[width=0.47\columnwidth]{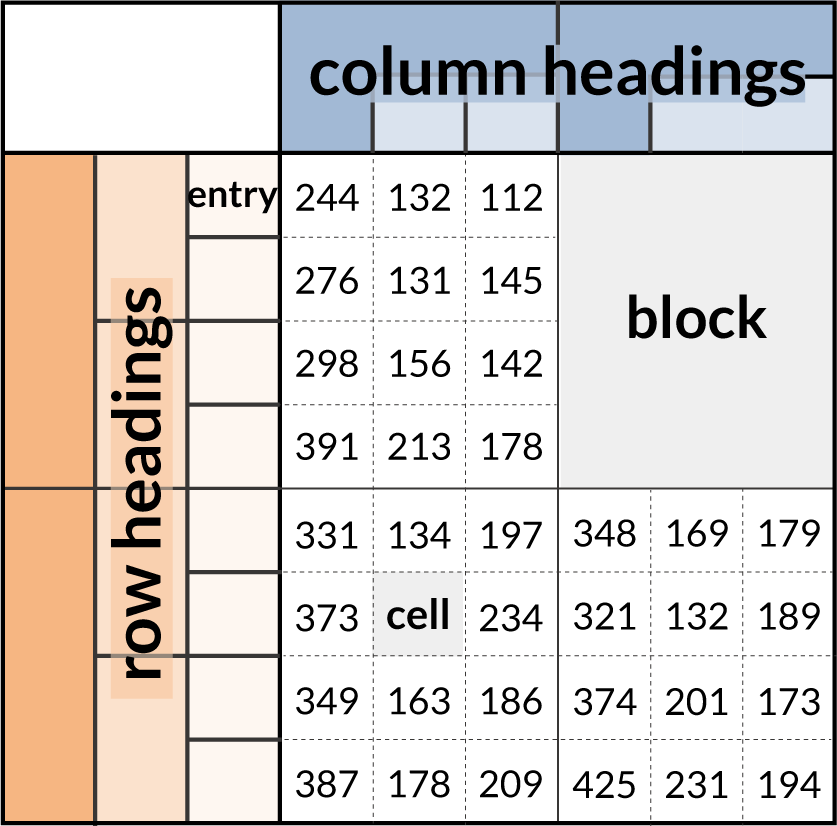}
    \label{fig:tablevis-example-b}
}\hfil
\subfloat[]{
    \includegraphics[width=0.47\columnwidth]{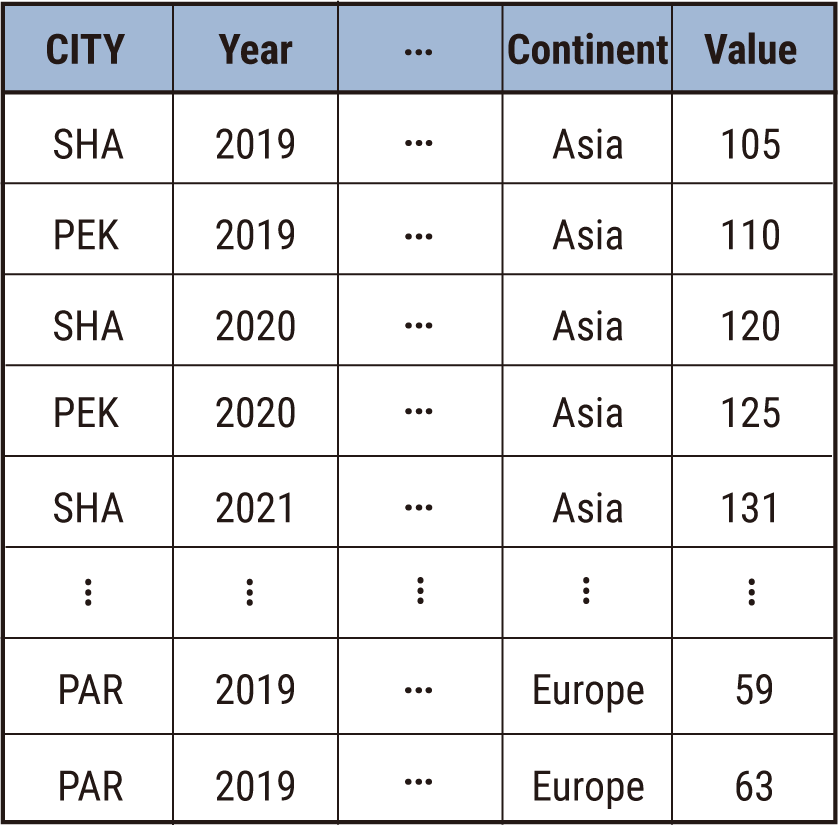}
    \label{fig:tablevis-example-c}
}\hfil
\subfloat[]{
    \includegraphics[width=0.47\columnwidth]{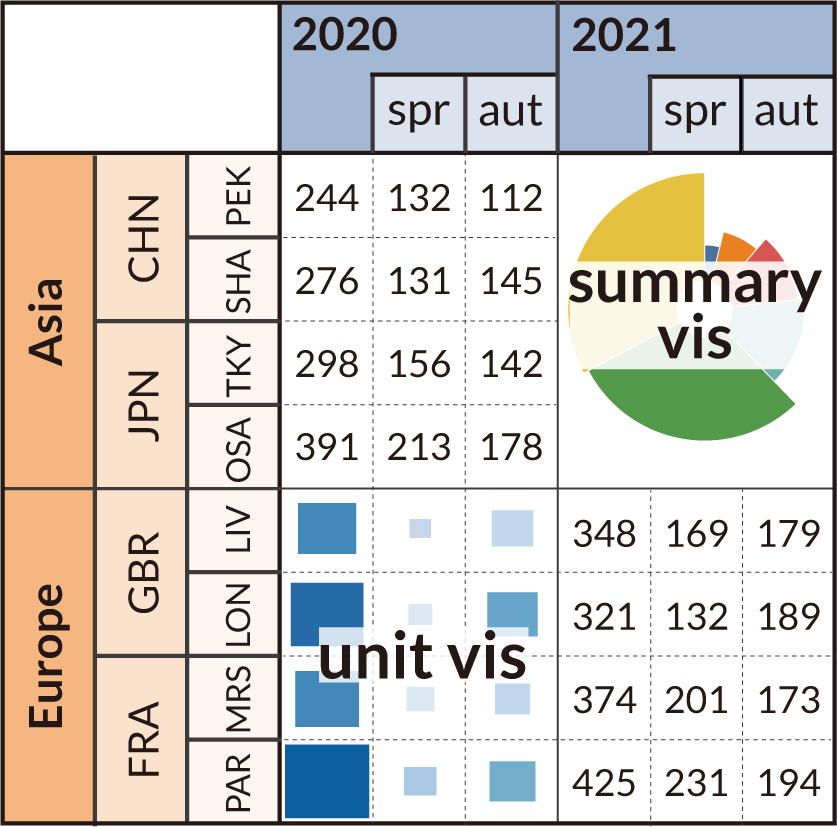}
    \label{fig:tablevis-example-d}
}\hfil
\caption{(a) is an example of a hierarchical table, while (b) provides annotations for the key constituents of this hierarchical table, encompassing cells, blocks, column headings, and row headings. (c) displays the corresponding flat table counterpart of (a), where the contents of the hierarchical table are exclusively positioned within the final column (value). (d) shows a hierarchical table visualization. This visualization involves the incorporation of summary visualizations within a contiguous block, alongside unit visualizations within individual cells.}
\label{fig:table-vis-example}
\end{figure}

\IEEEPARstart{T}{abular} data is a ubiquitous format that plays a prominent role in diverse applications~\cite{Estimating2005Scaffidi, 2018-Expandable-Dou, 2019-taggle-iv}. For example, financial analysts employ tabular data to record transactions and researcher scientists frequently utilize tabular data for the storage of experimental outcomes~\cite{yang2021numerical, yang2022TupleExtraction}. 
Accordingly, tables are widely used to represent tabular format data because they can organize and convey information clearly and intuitively. 
The pervasiveness and significance of tabular data make the table a vital approach for data organization and analysis across various disciplines. 
In particular, hierarchical tables~\cite{du2021tabularnet, li2022hitailor} are created with multi-level headers in contrast to flat tables, as shown in Fig.~\ref{fig:tablevis-example-a}. 
The hierarchical structure~\cite{schulz2011treevis, li2020GoTree, li2020BarcodeTree} of table headers enables the data organization in a compact approach for easy look-up and side-by-side comparison~\cite{wang2021tuta, cheng2022hitab, 2024-coinsight-li}.

The tabular technique~\cite{2019-taggle-iv} of table visualizations preserves the tabular layout while embedding visual representations within the underlying table to avoid additional cognitive load by splitting users' attentions~\cite{2006-integrating-ginns, 2006-AyersSweller, 2005-implications-sweller, 1998-cognitive-sweller}.
As illustrated in Fig.~\ref{fig:tablevis-example-b}, the tabular technique generally employs two approaches: visualizing individual data items using a single visual element~\cite{2018-insightable-atom} within each cell~\cite{2013-Lineup-Gratzl, 2016-heatmap-gu, 2014-Revisiting-charles, 1994-table-rao} or integrating summary visualizations to reveal patterns within a \textit{block} comprising a rectangular grouping of contiguous cells~\cite{2019-taggle-iv, 2011-visbrick}.
Through preserving the tabular layout, visualization results of tabular techniques can serve as users' navigational aids for exploring original datasets and comprehending the outcomes of analysis.

One existing study, named HiTailor~\cite{li2022hitailor}, has specified the manual process of embedding visual representations within hierarchical tables, involving table transformation, subset selection, and visualization of the selected data. 
Such a process is tedious and time-consuming. 
Moreover, the manually constructed hierarchical table visualizations may not necessarily yield rich insights into the underlying data.
Efficiently constructing hierarchical table visualizations that uncover substantial insights is a challenging task.
First, the creation of hierarchical table visualizations entails multiple steps, each requiring a multitude of decision-making processes. Consequently, the design space for such visualizations is vast. 
Second, the underlying data for hierarchical table visualizations is often intricate and multifaceted, encompassing many variables and relationships necessitating consideration. 
Such complexity presents significant challenges when endeavoring to generate insightful and meaningful interpretations of the data.

It is true that all hierarchical tables can be converted into flat tables, such as the hierarchical table depicted in Fig.~\ref{fig:tablevis-example-a} can be transformed into a flat table (Fig.~\ref{fig:tablevis-example-c}). However, the flat table arranges the entire table content of a hierarchical table within a single column, which significantly constrains the diversity of data selection and potential embedded visualizations. The headings within hierarchical tables achieve a more efficient data organization, thereby supporting the incorporation of a range of visualization outcomes. Moreover, the distinct organization of data within the headings of hierarchical tables also determines cells that correspond to specific entries, contributing to the variety of data block selections within hierarchical tabular data.

Previous research on the visual exploration and analysis of tabular data can be categorized into two distinct groups.
The first category pertains to defining insights within the tabular data~\cite{tang2017extracting, ding2019quickinsights, ma2021metainsight, lin2018bigin4}, such as the identification of outliers and trends. The second category revolves around recommending visualizations based on these insights, encompassing both single visualizations~\cite{demiralp2017foresight, peng2021dataprep} and sets of related visualizations~\cite{wang2019datashot, shi2020calliope, deng2022dashbot, bar2020automatically}. Nevertheless, these categories primarily focus on flat tables and generate independent visualization results. As such, they are not directly applicable to our scenario of embedding visualizations into hierarchical tables.

In this paper, we propose InsigHTable, a mixed-initiative system for realizing \textbf{Insig}ht-driven \textbf{H}ierarchical \textbf{Table} visualization construction. 
To accomplish this, we first model the construction process of hierarchical table visualizations as a Markov Decision Process and introduce a deep reinforcement learning (DRL) framework that integrates an auxiliary reward mechanism. 
To tackle the unique challenges associated with hierarchical table visualizations, we first summarize the data insights from hierarchical tables that are relevant either within a single block or across multiple related blocks, and define the number of data insights as the reward of the DRL framework.

The DRL framework comprises two modules: feature extraction module and action generation module.
The feature extraction module employs Graph Convolutional Network (GCN) to model the hierarchical structure of the table headings and Multi-Layer Perceptron (MLP) as well as Bidirectional Long Short-Term Memory (Bi-LSTM) to represent the data content. 
The action generation module encompasses two stages, table transformation and visualization embedding.
Considering no rewards provided in the table transformation stage, we utilize an auxiliary reward to stimulate agent exploration, and employ a joint optimization method to balance the extrinsic and auxiliary rewards.
The agent within the DRL framework takes a series of actions to maximize rewards associated with discovering data insights. 
Through continuous interaction with the environment, the agent optimizes its decision-making process of creating hierarchical table visualizations from feedback, enhancing users' capacity to discover more insights from hierarchical tables.

We conduct two case studies and quantitative experiments, encompassing the evaluation of hyperparameters, ablation studies, robustness testing, and comparisons with heuristic methods. All experiments are conducted on real-world datasets.
The case studies demonstrate that InsigHTable can help data analysts  understand and explore hierarchical tables effectively.
In addition, the results of quantitative experiments show the performance and superiority of the DRL framework utilized in InsigHTable.

In summary, the main contributions are as follows:
\begin{itemize}[leftmargin=3.5mm]
    \item An insight formulation tailored to hierarchical tables, considering data insight within a single block and across multiple blocks.
    \item A DRL framework to realize the mixed-initiative construction of hierarchical table visualizations, which incorporates auxiliary reward mechanisms to enhance the agent's exploratory capabilities.
    \item The InsigHTable prototype system that assists users in analyzing hierarchical tables and obtaining data insights.
\end{itemize}

\section{Related Work}
\label{sec:related-work}
 
This section reviews the literature on automatic insight discovery of tabular data and reinforcement learning techniques for visualizations.  
\subsection{Automatic Insight Discovery of Tabular Data}
Data insights, \textit{i.e.}, interesting data patterns, are critical in analyzing data and play a fundamental role in creating effective visualizations. 
Several studies have concentrated on extracting data insights, which can aid users in gaining a deeper understanding of the data.
Top-k insights~\cite{tang2017extracting} focuses on identifying outliers and trends in multidimensional data. 
Microsoft Power BI's QuickInsights~\cite{ding2019quickinsights} framework provides a complete system for insight discovery and representation, supporting various types of insights.  
Furthermore, MetaInsight~\cite{ma2021metainsight} proposes a novel scoring function to quantify the usefulness of insights. 
BigIn4~\cite{lin2018bigin4} combines data cubes and the Approximate Query Processing technique to provide interactive insight recommendations for a large-scale dataset. Additionally, DataSite~\cite{cui2019datasite} is a web-based platform that enables automatic analysis and calculation while users visualize and analyze data.

Recommendation systems for visualizations have been a popular topic in several studies.
Some of these studies focus on recommending visualizations for individual charts based on insights extracted from data. 
For example, Foresight~\cite{demiralp2017foresight} uses insight indicators to recommend visualization, while DataPrep.EDA~\cite{peng2021dataprep} provides programming interfaces for this functionality in Python.
In contrast, some studies have focused on the significance of recommending multiple related visualizations rather than one single chart.
Datashot~\cite{wang2019datashot}, for instance, automatically generates fact sheets for tabular data using extracted insights. 
Calliope~\cite{shi2020calliope} builds on Datashot's automatic insight generation technique to generate data stories in a logical sequence to enhance users' comprehension. 
In a different approach, Voder~\cite{srinivasan2018augmenting} combines the natural language generation technique with insight calculation to derive corresponding data facts, aiming to enhance data comprehension by presenting insights in a more accessible format.
Together, these studies offer various ways to explore and visualize insights in data, highlighting the importance of considering multiple visualizations and techniques for practical data analysis.

Previous research has primarily focused on analyzing and visualizing simple flat tables rather than hierarchical tables. Some studies have explored interactive approaches for constructing hierarchical table visualizations~\cite{li2022hitailor} and collected hierarchical tables~\cite{cheng2022hitab}, but did not investigate the data insight extraction. 
Additionally, existing research has predominantly concentrated on generating independent visualizations rather than embedding visualizations into tabular data, which often require several steps about data transformations and visualizations. 
Our work addresses this gap by exploring techniques for extracting insights from hierarchical tabular data and efficiently constructing table visualizations.

\subsection{Reinforcement Learning for Visualizations}
Reinforcement learning is a machine learning approach that allows agents to learn by interacting with an environment through trial and error, which can be categorized into two main branches: classical reinforcement learning and DRL. Classical reinforcement learning methods, such as Q-learning~\cite{watkins1992q} and SARSA, are based on dynamic programming.
DRL algorithms, such as Deep Q-Networks~\cite{mnih2015human} and Policy Gradient~\cite{sutton1999policy}, incorporate deep neural networks to handle high-dimensional state and action space and have achieved impressive results in various domains including Go~\cite{silver2016mastering}, real-time strategy games~\cite{vinyals2019grandmaster}, and nuclear fusion~\cite{degrave2022magnetic}.

Reinforcement learning techniques have been applied in some visualization research.
Some studies~\cite{wu2020mobilevisfixer, chen2021vizlinter, el2017progressive, kassel2019online} use classical reinforcement learning, which does not incorporate deep neural networks, to tackle problems with limited state function mappings.
For instance, MobileVisFixer~\cite{wu2020mobilevisfixer} employs a reinforcement learning framework to automate the redesign of mobile-friendly visualizations, while Vizlinter~\cite{chen2021vizlinter} helps users detect and correct defects in existing visualizations.
In contrast, El-Assady et al.~\cite{el2017progressive} proposed a modularized visualization analysis framework that leverages a user-driven reinforcement learning process to address the interpretability and adaptability issues of topic models. 
While these approaches focus on optimizing and revising visualizations, classical reinforcement learning methods can provide effective solutions within a limited state space and action space. 

Comparing to classical reinforcement learning, DRL holds the potential to address more intricate visualization problems.
To handle high-dimensional input data, such as images and tables, some studies~\cite{hu2021shape, zhou2021table2charts, tang2020plotthread, deng2022dashbot, bar2020automatically} employ deep neural networks that enable the agent to learn state abstractions and policy approximations directly from the input data.
Hu et al.~\cite{hu2021shape} utilized the  DRL~\cite{bello2016neural} and pointer network~\cite{vinyals2015pointer} algorithm to address the coordinate ordering problem in star-glyph sets.
Additionally, several research efforts have focused on visualization generation and recommendation.
For example, Table2Charts~\cite{zhou2021table2charts} employs deep Q-learning with a copy mechanism and heuristic search to generate table-to-chart sequences following a chart template.
Similarly, DashBot~\cite{deng2022dashbot} generates analytical dashboards that leverage established visualization knowledge and the estimation capacity of DRL.
ATENA~\cite{bar2020automatically} models exploratory data analysis as a control problem and designed a novel DRL architecture to optimize the generation of analysis notebooks.
The latest Visail~\cite{shi2023supporting} can support the exploratory visual analysis of time-series data, generating exploratory visual analysis sequences consisting of multiple steps.
In addition to chart generation and recommendation, PlotThread~\cite{tang2020plotthread} uses a DRL framework to train an agent that assists users in exploring the design space of storyline visualizations and generating well-optimized results.

While many studies have employed reinforcement learning to address challenges in the visualization community, these applications encounter difficulties in constructing hierarchical table visualizations. 
This is attributed to the process's two distinct stages, and one stage is characterized by sparse rewards.
In response to this challenge, this work employs a novel reinforcement learning framework with two distinct stages, incorporating auxiliary rewards to address the issue.

\section{PRELIMINARIES}
This section introduces the background of reinforcement learning techniques utilized in this work and then formulates the task of constructing hierarchical table visualizations as a sequential decision-making problem. 
\subsection{Background on Reinforcement Learning}
Reinforcement learning explores how intelligent agents can learn to maximize rewards in uncertain environments, which involves both an agent and an environment.
The agent is a deep neural network that learns a target task, while the environment is the scene in which the agent operates. Throughout the reinforcement learning process, the agent and the environment continually interact. 
When the agent obtains a certain state in the environment, it outputs an action, also known as a decision. This action is then executed in the environment, and the environment outputs the next state and the reward associated with the action taken by the agent. The reward is a scalar feedback signal given by the environment, which indicates how well the agent performed a specific strategy in a given step.
The purpose of the agent is to obtain as much reward as possible from the environment. The formula for calculating the cumulative reward $R_t$ is as follows:
$$R_t = r_t + \gamma r_{t+1} + \gamma^2 r_{t+2} + \cdots = \sum_{k=0}^{\infty} \gamma^k r_{t+k}$$  
where $\gamma$ represents the discount factor that controls the weight of future rewards during the calculation of the $R_t$. 
A value of $\gamma \in [0, 1]$ is commonly used, where a value of $\gamma$ close to 1 indicates that future rewards have more influence.

\subsection{Problem Formulation}
A hierarchical table is a common data format that often requires manual exploration to discover data insights. 
The construction of hierarchical table visualizations is a time-consuming and laborious process. 
InsigHTable is designed to reduce the efforts of human operations for visualization construction. 
We define the operations proposed in HiTailor~\cite{li2022hitailor} for interactive hierarchical table transformation, including \textit{transpose}, \textit{tolinear}, and \textit{swap}, as actions of the agent in the DRL framework.
After table transformations and visualization embedding, it is required to measure the quality of hierarchical table visualizations.
Previous work on insight extraction~\cite{tang2017extracting, ding2019quickinsights} and the design principles~\cite{deng2022dashbot} inspire us to evaluate the entire table transformation and visualization for hierarchical tables.

Insight-driven hierarchical table transformation and visualization consist of two challenges.
First, hierarchical tables are complex because of the presence of multi-level headings, and each arrangement of table headings corresponds to a unique table layout. 
Second, the operations of table heading transformation and selection influence the visualization process, making the hierarchical table visualization construction a sequential decision-making problem.
Furthermore, selecting data from hierarchical tables needs to determine the row and column headings, and the multi-level characteristic of table headings also results in numerous possibilities.

\section{METHODS}
\subsection{Insight Formulation}
\label{sec:insight}
\begin{figure*}[!h]
    \centering
    \includegraphics[width=\textwidth]{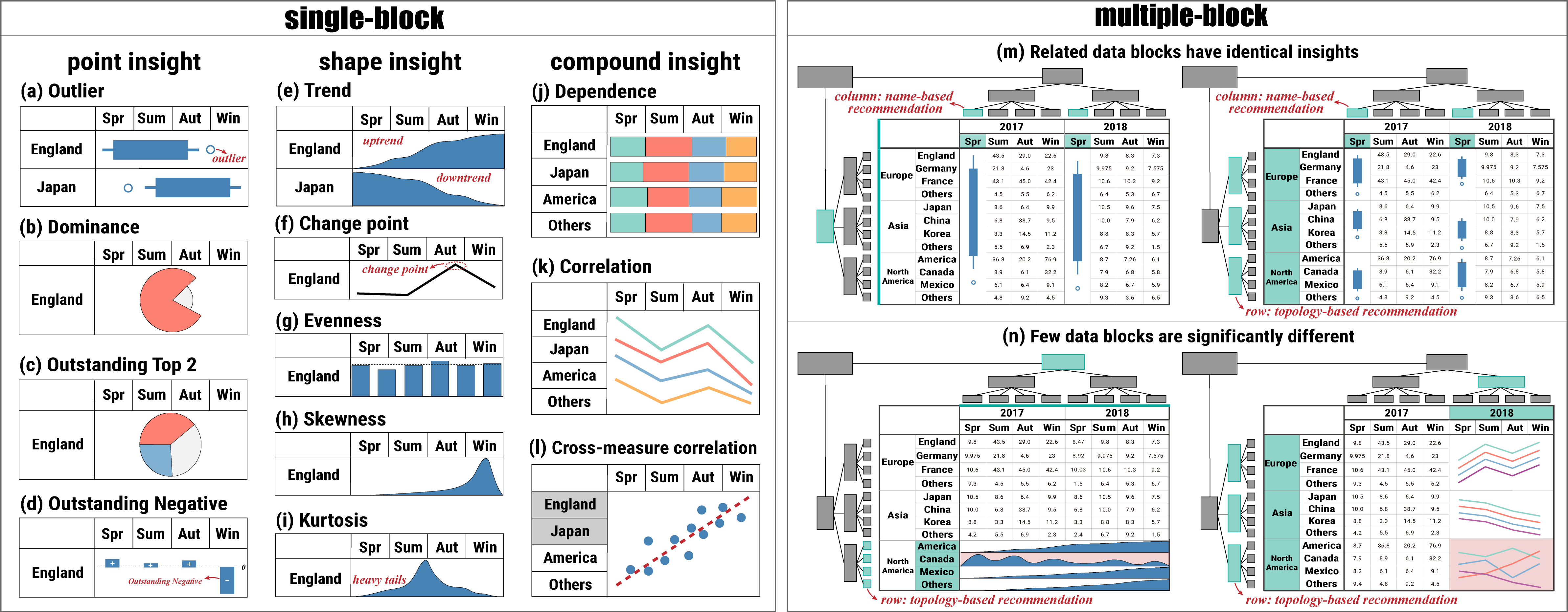}
    \caption{Insight classifications for hierarchical tabular data. The left part shows the single-block insights, encompassing point insights, shape insights, and compound insights. The right part shows the multiple-block insights, which is defined through both name-based recommendation and topology-based recommendation mechanisms. The selected entries within the table headings are highlighted in green. }
    \label{fig:insight}
\end{figure*}

The insight formulation of this paper is grounded by the existing studies for insights within flat tables and primitive analytical tasks.
Previous studies, exemplified by QuickInsights~\cite{ding2019quickinsights}, Top-k insights~\cite{tang2017extracting}, and DataShot~\cite{wang2019datashot}, have formulated metrics for extracting insights in flat tables.
The above work defines the data subspace for extracting data insights and then visualizes the extracted data insights independently from the original table. 
Some studies~\cite{amar2005low, 2013-designspace-Schulz, 2004-taxonomy-tory} also summarize primitive analytical tasks in information visualization, which serve as a common language for information visualization systems and provide the groundwork for developing insight taxonomy.
In particular, the insight formulation in this work is tailored to integrate visual representations directly into hierarchical tables. 
Therefore, it is imperative to ensure that the data subspace comprises continuous table cells under the current hierarchical table arrangement.
The arrangement of multi-level headings in hierarchical tables dictates the corresponding data content.
In this context, HiTailor~\cite{li2022hitailor} defines a block as a cluster of cells in hierarchical tables uniquely determined by row and column entries.
Consequently, we adopt the data block as the primary unit for insight extraction and computation.

The extraction and calculation of insight in hierarchical tables can be divided into two types. 
The first type is based on a single data block. A data block can be treated as a data subspace~\cite{tang2017extracting}, enabling the use of existing methods for insight extraction. 
However, unlike flat tables, hierarchical table headers exhibit a multi-level structure, introducing correlations between different data blocks. 
Consequently, for the task of insight extraction in hierarchical tables, we account for the relations between data blocks, including name-based and topology-based relations. 
Details of these relations are explained in Sec.~\ref{subsection:multiple-block-insight}.
This consideration extends the insight extraction from a single block to multiple blocks.

\subsubsection{Single-block Insight}
\label{sec:single-block-insight}

The data insights derived from a single block are interesting data patterns calculated from these data items.
Our consolidation of single-block insights draws from established works such as QuickInsights~\cite{ding2019quickinsights},  Foresight~\cite{demiralp2017foresight}, and Top-k insights~\cite{tang2017extracting}. 
These insights are classified into three categories: \textbf{point insight}, \textbf{shape insight}, and \textbf{compound insight}.

\textbf{Point insight} refers to the characteristic of having several prominent data values within a data block.
It consists of four insight types, outlier, dominance, outstanding top two, and outstanding negative.

(1) \textbf{Outlier} indicates that data items are significantly different from the majority within the data block~\cite{ding2019quickinsights, wang2019datashot}.
This insight corresponds to the \textit{Finding Anomalies} task within a given set of data cases~\cite{amar2005low}, which is significant and often provides a basis for further explorations.
We employ two methods for detecting outliers. 
The first method uses interquartile ranges to determine the upper and lower bounds~\cite{srinivasan2018augmenting}. 
More specifically, we consider the data points that fall outside of the range defined by $Q3+3 \times IQR$ or $Q1-3 \times IQR$ as outliers. 
As shown in Fig.~\ref{fig:insight}(a), we visualize outliers using a box plot. 

The other approach is to fit a power-law function to the data and then calculate the probability of a value being significantly larger than other values based on the predicted error~\cite{tang2017extracting}.
These outliers can be visualized using a bar chart.

(2) \textbf{Dominance} refers to the degree of superiority that one random variable has over one or more others~\cite{ding2019quickinsights}. 
In other words, one variable has a significant advantage over the others (\textit{e.g.}, at least 50\%), which is related to the \textit{Finding Extremum} task~\cite{amar2005low}.
We use the pie chart or the radial plot to visualize the dominance insight, as shown in Fig.~\ref{fig:insight}(b).

(3) \textbf{Outstanding Top Two} refers to two variables having an advantage over the others (\textit{e.g.}, both $\geq$ 34\%)~\cite{ding2019quickinsights}. 
The top 2 insight can also be visualized using a pie chart or a radial plot, as shown in Fig.~\ref{fig:insight}(c).

(4) \textbf{Outstanding Negative} refers to the pattern where the most negative value is significantly lower than all remaining values~\cite{ding2019quickinsights}. We employ the bar chart to visualize this insight, as shown in Fig.~\ref{fig:insight}(d).

\textbf{Shape insight} characterizes the general trends or distributions within a specific data block and consists of five insight types: trend, change point, evenness, skewness, and kurtosis.

(1) \textbf{Trend} refers to the presence of a consistent upward or downward pattern in the data items, especially within the context of time series data~\cite{tang2017extracting, ding2019quickinsights}. 
We can employ the following formula to detect trends: \\
\centerline{$trend(b) = r^2 \cdot (1-p)$} \\
where $r^2$ is the coefficient of determination and $p$ is the $p$-value that indicates the likelihood of observing slope values equal to or greater than the slope of the ascending or descending trend. 
The above formula is applicable for both upward and downward trends.
This insight corresponds to the \textit{Finding Temporal Relations} task from the target perspective~\cite{2013-designspace-Schulz}.
We can visualize trends using either  a horizon graph or a line chart, as shown in Fig.~\ref{fig:insight}(e).

(2) \textbf{Change Point} refers to identifying a notable shift or transition in the behavior or values of a random variable within a data block~\cite{ding2019quickinsights}. 
This insight involves pinpointing a specific point at which the variable's behavior or characteristics noticeably deviate from its previous pattern.
Such a deviation often suggests a change in the underlying data-generating process or external conditions influencing the variable.
The change point insight can provide valuable information about critical shifts in trends, distributions, or relationships within the data.
We can visualize the change point insight using a line chart or a horizon graph, as shown in Fig.~\ref{fig:insight}(f).

(3) \textbf{Evenness} refers to the degree of similarity among variables and can be measured using the coefficient of variation~\cite{ding2019quickinsights}, which is calculated by dividing the standard deviation $\sigma$ of the variable by its mean $\mu$.
A lower coefficient of variation indicates a more uniform distribution of variables. We employ a bar chart for visualizing evenness, as shown in Fig.~\ref{fig:insight}(g).

(4) \textbf{Skewness} measures the degree of asymmetry in the distribution of values~\cite{demiralp2017foresight}, with larger absolute values indicating greater deviation from symmetry.
The following formula can be applied to detect skewness:\\
\centerline{$\kappa_1(b)=n^{-1} \sum_i^n\left(b_i-\mu_b\right)^3 / \sigma_b^3$}, \\
where $\mu_b$ is the mean of the data, $\sigma_b$ is the standard deviation of the data, $n$ is the number of data points, and $b_i$ is the $i^{th}$ data point. We can visualize skewness using a density plot, as shown in Fig.~\ref{fig:insight}(h).

(5) \textbf{Kurtosis} measures the degree of peakedness of the value distribution within the data block, with a larger absolute value indicating a sharper distribution~\cite{demiralp2017foresight}.
The following formula can be employed to calculate kurtosis:\\
\centerline{$\kappa_2(b)=n^{-1} \sum_i^n\left(b_i-\mu_b\right)^4 / \sigma_b^4$}, \\
We also visualize the kurtosis insight using a density plot, as shown in Fig.~\ref{fig:insight}(i).

The above three insights (Evenness, Skewness, and Kurtosis) are related to the \textit{Characterizing Distribution} task in the low-level task taxonomy~\cite{amar2005low}.

\textbf{Compound Insight} refers to the insights with different groups of data items within a data block and consists of three insight types: dependence, correlation, and cross-measure correlation.

(1) \textbf{Dependence} insight involves dividing the data set by the row or column attributes, treating the row and column of the blocks as two categorical variables~\cite{demiralp2017foresight}, and verifying their independence using the Chi-square test by calculating the $p$-value. 
The results of the chi-square test are visualized using the normalized stacked bar chart in Fig.~\ref{fig:insight}(j). 

(2) \textbf{Correlation} insight also divides the data set into different groups. Unlike the dependence insight, it is specifically designed to verify the pairwise correlations between different groups~\cite{demiralp2017foresight, ding2019quickinsights, cui2019datasite, wang2019datashot}.
When most groups exhibit a significant correlation ($p$-value \textless 0.05), we consider this block with multiple rows to consist of a correlation insight. We utilize the multi-series line chart to visualize the correlation insight, as shown in Fig.~\ref{fig:insight}(k).

(3) \textbf{Cross-measure correlation} insight defines the relationships between two measures~\cite{ding2019quickinsights} using the magnitude of the Pearson correlation coefficient $\vert \rho(x, y)\vert$, where $\rho(x, y)$ is equals to ${\textstyle \sum_{i=1}^{n}({x_i - \mu_x})(y_i-\mu_y)}/(\sigma_x \sigma_y)$. This insight is visualized using a scatter plot visualization approach, as shown in Fig.~\ref{fig:insight}(l). 

The above three insights (Dependence, Correlation, and Cross-measure correlation) are related to the \textit{Correlate} task in the low-level task taxonomy~\cite{amar2005low}.

\subsubsection{Multiple-block Insight}
\label{subsection:multiple-block-insight}

Multiple-block insights are computed across data blocks, with users selecting a specific data block and determining other related data blocks through the name-based and topology-based mechanisms.
As shown in the right part of Fig.~\ref{fig:insight}, a data block is jointly determined by two entries in hierarchical tables' row and column headings. For related data blocks, one entry is the same, and the other entries have the same name or are topologically related, such as parent-child or sibling relations.

The computation of multiple-block insights is grounded in single-block insights, introduced in Sec.~\ref{sec:single-block-insight}. 
When recommended data blocks share the same structure, high-level insights can be extracted by comparing insights across different related blocks.
The right part of Fig.~\ref{fig:insight} illustrates these two types of insights for multiple blocks. 
The first type, as shown in Fig.~\ref{fig:insight}(m), indicates that all data blocks contain identical insight. 
The second type, as shown in Fig.~\ref{fig:insight}(n), is that the insight of one data block significantly differs from the others among related data blocks.
The resulting visualizations are embedded in all recommended blocks, with the red background color indicating the data block that does not contain the specific data insight.

\subsection{Markov Decision Process}
\label{sec:markow-decision-process}
The construction of hierarchical table visualizations involves a sequential decision-making process, where each state depends only on the previous state. Therefore, we formalize the transformation and visualization of hierarchical tables as a Markov Decision Process~\cite{bellman1957markovian}, defined as a tuple $(S, A, T, R, \gamma)$, consisting of a state space $S$, an action space $A$, a transition function $T$, a reward function $R$, and a discount factor $\gamma$. 
The transition function $T$ defines the probability of moving from one state to another given an action. 

By formulating the problem as a Markov Decision Process, we propose a DRL framework to learn an optimal policy for hierarchical table transformation and visualization.
Note that the DRL framework only selects the data blocks to be embedded with visualizations by providing operations for table transformation and table heading selection, and then heuristic methods are used to extract different types of insights and determine the corresponding visualization.

\subsubsection{State Space}
The state space $S$ includes all possible states of the hierarchical table heading's structure and their corresponding tabular data and is defined as follows,
$$S\stackrel{\Delta}{=} \left\{s_t = (H_t, D_t)\right\}$$
where $H_t$ represents the hierarchical structure of table headings at the $t$-th time step, which contains multiple levels of row and column headings, and 
$D_t$ represents the numerical content of tabular data, along with a specialized mask indicating whether the cells are embedded with visualizations.

\subsubsection{Action Space}
\label{sec:action-space}
\begin{figure}[!h]
    \centering
    \includegraphics[width=\linewidth]{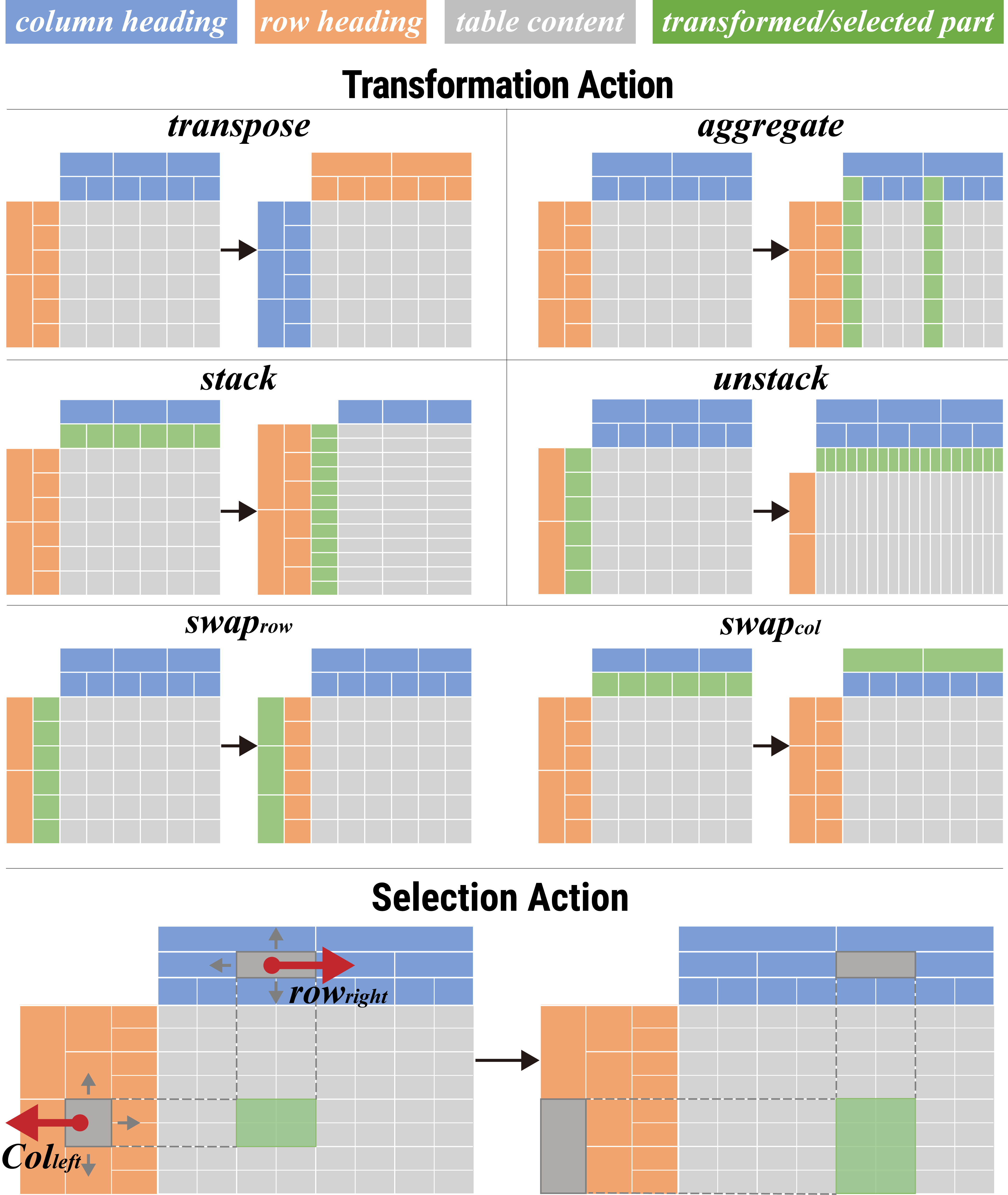}
    \caption{The actions conducted by the agents of the reinforcement learning model are divided into two groups, the \textit{transformation} action of the hierarchical table, and the \textit{selection} action of data items to be visualized.}
    \label{fig:action}
\end{figure}

Considering the headings of hierarchical tables determine how the inner data items are organized, we define the action space $A$ based on the table headings, as shown in Fig.~\ref{fig:action}, where each action
$a_t\in\{$\textit{transpose}, \textit{aggregate}, \textit{stack}, \textit{unstack}, 
\textit{swap}$_{\textrm{row}}$, 
\textit{swap}$_{\textrm{col}}$, 
\textit{row}$_{\textrm{up}}$, \textit{row}$_{\textrm{down}}$, \textit{row}$_{\textrm{left}}$, \textit{row}$_{\textrm{right}}$, \textit{col}$_{\textrm{up}}$,  \textit{col}$_{\textrm{down}}$, \textit{col}$_{\textrm{left}}$, \textit{col}$_{\textrm{right}}\}$.
Furthermore, we divide the actions into two categories: transformation and selection. The transformation type in action space consists of five actions, shown in Fig.~\ref{fig:action}. 
The transpose action exchanges the row and column headings of the entire hierarchical table. The aggregate action is designed to derive some values based on multiple rows or columns under a specific category, such as sum and average. The stack action moves the innermost level in the column heading to the innermost level in the row heading. 
Conversely, the unstack action moves the innermost row heading to the innermost column heading. The swap action exchanges two levels in the row or column headings, where users specify which two levels to swap. Based on these actions, users can transform the heading structures and select specific data items flexibly. However, some transformation actions may result in empty cells in the hierarchical table because the corresponding data items for these heading combinations do not exist. The transformation process still continues in such cases, but visualization actions cannot be performed.

The \textit{selection} type of actions is designed to locate a specific data block within hierarchical tabular data by manipulating the selected entries of row and column headings. These actions involve moving the selected entry, which is initially positioned at the top-left corner. The actions for both row and column headings are identical and include four directions: up ($row_{\textrm{up}}$), down ($row_{\textrm{down}}$), left ($row_{\textrm{left}}$), and right ($row_{\textrm{right}}$). As shown in Fig.~\ref{fig:action}, the selected entry in the row and column headings uniquely determines a data block that can be modified by moving the entries. For the discovered insights during the selection process, the agent utilizes the corresponding visualizations to present the data blocks.

In particular, our framework prohibits the repeated visualization of data items from the same cells to avoid the overlap of embedded visualizations. 
As a result, each action generated by the agent undergoes validation, preventing the selection of the same regions for embedding visualizations. 
However, over multiple iterations of the hierarchical table visualization construction process, data items from the same region may be visualized using diverse charts. 
This approach promotes diversity in visualization results rather than constraining them to a fixed visualization outcome.

\subsubsection{Reward Function}
\label{sec:reward-function}
The reward function is pivotal in shaping the learning process for the agent.
In defining the reward function, we have established three metrics---Area Ratio, Insight Ratio, and Evenness Ratio---to evaluate the quality of the eventual discovery of data insights.

\textbf{Area Ratio (\textit{AR}).}
After the agent completes the visualization generation, we calculate the percentage of data cells in the agent's visualization:
$$AR = \frac{A_{v}}{A_{d}}$$ where $A_{v}$ denotes the number of cells within the region that has been visualized by the agent, and $A_{d}$ denotes the number of cells in the whole table.

\textbf{Insight Ratio (\textit{IR}).}
We have defined a scale to represent the percentage of the total insight types that have been discovered by the agent:
$$IR = \frac{N_d}{N_t}$$
where $N_d$ is the number of discovered insight types, $N_t$ is the total number of insight types.

\textbf{Evenness Ratio (\textit{ER}).} 
To balance the impact of various insights, we calculate the Shannon entropy based on the specific region of the visualization for each type of insight, aiming to balance the quantity of insights and produce a diverse visualization result:
$$ER = -\sum_{i=1}^{N_t}p_i\ln(p_i)$$
where $N_t$ is the number of insight types discovered and $p_i$ represents the probability of each insight covering a certain number of cells.

The reward function $R$ evaluates the quality of the hierarchical tabular visualization after an action is performed. 
To find more insights as defined in Sec.~\ref{sec:insight}, we define $r^c_t$ as follows:
\begin{align*}
\begin{split}
r^c_t &= \eta_1 \Delta AR_t \\
\Delta AR_t &= AR_t - AR_{t-1}
\end{split}
\end{align*}
where $\eta_1$ indicates the weight and $\Delta AR_t$ denotes the ratio of cells containing insights that were discovered and visualized at the $t$-th time step. 
This reward serves as motivation for the agent to actively seek out and discover more insights.

To encourage the exploration of diverse insights, we propose the use of diversity rewards denoted as $r^d_t$:
\begin{align*}
\begin{split}
r^d_t &= \eta_2 \Delta IR_t + \eta_3 \Delta ER_t \\
\Delta IR_t &= IR_t - IR_{t-1} \\
\Delta ER_t &= ER_t - ER_{t-1}
\end{split}
\end{align*}
where $IR_t$ and $ER_t$ denote the insight ratio and evenness ratio at the $t$-th time step.
The diversity reward is a metric utilized to evaluate the diversity of generated results, with a higher value indicating the discovery of multiple insights. 

By combining $r^c_t$ and $r^d_t$ defined above, we define the reward as:
\begin{align*}
r_t &= r^c_t + r^d_t\\
 &= \eta_1 \Delta AR_t + \eta_2 \Delta IR_t + \eta_3 \Delta ER_t
\end{align*}
which represents the reward that the agent can receive from the environment at the $t$-th time step.

\subsection{Model Architecture}

\begin{figure*}[!h]
    \centering
    \includegraphics[width=\textwidth]{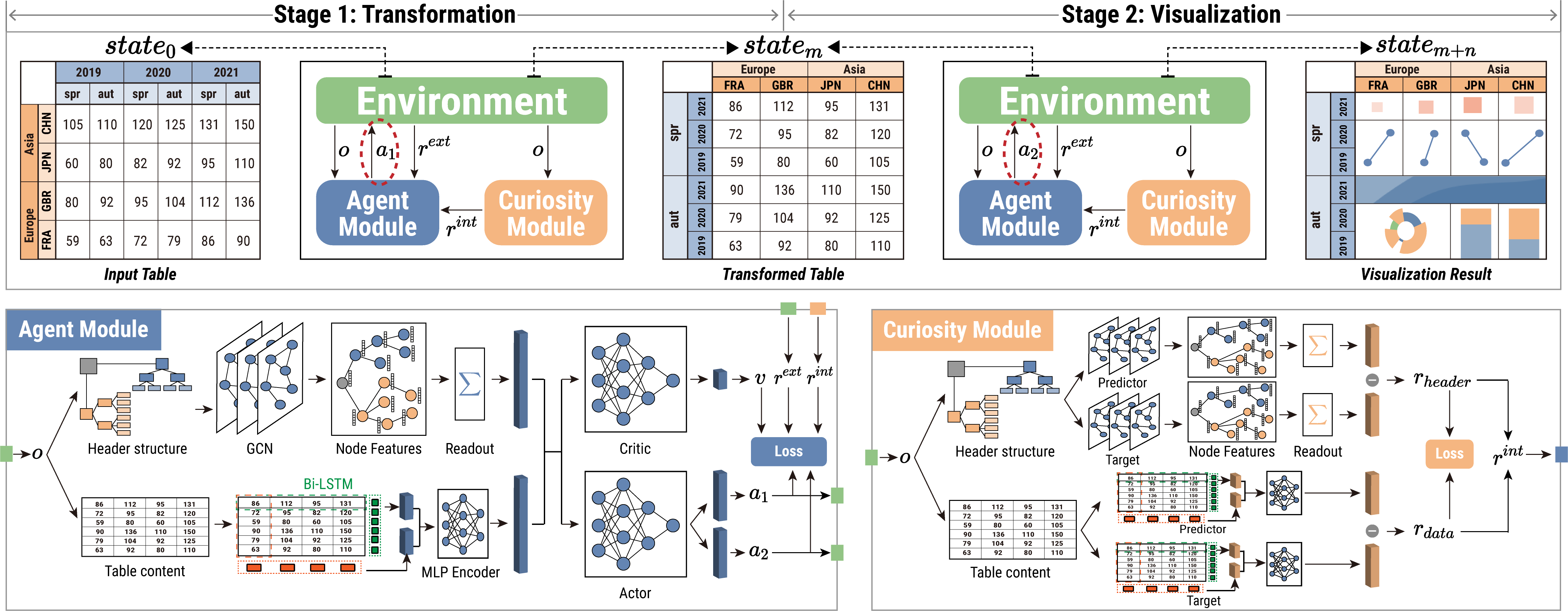}
    \caption{The construction process of InsigHTable involves two stages: transformation and visualization. These stages perform different operations on the hierarchical table. During the transformation stage, the table headers and content are updated accordingly. In the visualization stage, data blocks within the hierarchical table are identified, and visualization results are inserted. $state_0$ represents the input table, while $state_m$ denotes the transformation results. $state_{m+n}$ corresponds to the visualization results. The model's structure remains the same in both stages, except for the action output by the agent module, which is indicated by an ellipse with a dotted border to highlight their differences. The bottom part of the model structure consists of the agent module and the curiosity module.}
    \label{fig:model}
\end{figure*}

The DRL model comprises two modules: the agent module and the curiosity module. The agent module takes hierarchical tables as input and generates an action. The curiosity module is a supplementary component designed to encourage the agent to explore the novel states. 
In this section, we introduce the architecture of the deep neural network. First, we explain the two parts of the agent module: feature extraction and agent network architecture. Next, we introduce the curiosity module, especially the auxiliary reward mechanism.

\subsubsection{Feature Extraction}
\label{sec:feature_extraction}

The data representation module is a crucial part of deep neural networks. 
Efficient data representation can improve the training effectiveness of neural networks and achieve improved performance. 
Previous studies, such as VizML~\cite{hu2019vizml} and DashBot~\cite{deng2022dashbot}, have primarily focused on flat tables, addressing the challenge by computing numerous handcrafted features and inputting them into the network for visualization recommendations. 
However, representing hierarchical tables is a more challenging task. 
Considering the states of hierarchical table headings undergo continuous transformations, it is crucial to capture changes in the state of hierarchical table heading structures and the state of embedding visualizations.

In our study, we have crafted a model structure to tackle the challenge of representing hierarchical tables. 
Specifically, we partition the hierarchical table into two distinct parts, the hierarchical heading and the tabular data content. 
Subsequently, we represent the hierarchical heading structure and input it to the GCN~\cite{kipf2017semisupervised} Encoder.
As for the table content, we adopt a network design inspired by the TabularNet approach. 
Two Bi-LSTM layers are employed to capture the table content, and the resulting output is then passed to an MLP Encoder.

\textbf{The embedding of hierarchical table heading structure.} We adopt a tree-based representation approach~\cite{wang2021tuta} to model the heading structure of hierarchical tabular data, as shown in the \textit{heading structure} of Fig.~\ref{fig:model}.
Under this approach, each node in the tree is denoted as $x_i$ using its node index $i$. In contrast, the edges connecting the nodes represent the parent-child relationships between entries in the headings and are labeled as $e_{i,j}$, indicating that node $i$ is the parent of node $j$. 
Additionally, three virtual nodes are incorporated: the whole root node, the root node of row headings, and the root node of column headings. These nodes connect the hierarchical structures present in the row and column headings, thereby facilitating the integration of the two hierarchical structures.

In addition, the DRL framework employs a pre-trained BERT~\cite{kenton2019bert} model to obtain the content information of table headings, denoted as $ch_i$.
Additionally, the representation of hierarchical table headings reflects the current selection status of the visualization area using the selection status of the node denoted as $cs_i$, where $1$ denotes that the node is selected in the column heading, and $2$ denotes that the node is selected in the row heading, and $-1$ denotes that the node is not selected, indicating that the cells underlying this node are not involved in the visualization results. 

The feature representation $c_i$ for each node $x_i$ is obtained by contacting the semantic information $ch_i$ of the corresponding table heading and the header selection state $cs_i$.
The first-level node feature for each element in the heading $x_i$ is defined as follows:
$$
h_i^{(0)}=\operatorname{ReLU}\left(\operatorname{MLP}\left(c_i\right) \right)
$$
where ReLU is the activation function. 
Besides node features, we define edge features for each parent-child relationship $e_{i,j}$ as a one-hot encoding vector to obtain the adjacency matrix:
$$
w_{i, j}=\operatorname{OneHot}\left(e_{i, j}\right)
$$
After performing multiple rounds of graph convolution, the feature vector of node $x_i$ aggregates the features of each node and its neighbors based on the sum of their feature vectors and is defined as follows:
$$
h_i^{(l)}=\operatorname{MLP}\left(h_i^{(l-1)}+\sum_{j \in N(i)} w_{i, j} h_j^{(l-1)}\right)
$$
where $h_i^{(l)}$ is the feature vector of node $i$ at the $l$-th layer of the graph convolution, and $N(i)$ is the set of neighbors of node $i$.
The graph-level representation is obtained by performing a readout operation, where we choose to use the mean operation:
$$h_H=\frac{1}{N}\sum_{i=1}^N h_i^{(L)}$$
where $L$ is the number of graph convolution layers, and $h_H$ is the final representation of the entire heading structure.
Note that any transformation actions performed on a hierarchical table change the state of the headings. Consequently, the connectivity status of the edges in the table headings is affected, resulting in changes of the headings' features.

\textbf{The embedding of hierarchical table content.} In addition to embedding the heading structures, it is essential to represent the visualization embedding status and the table content as inputs to the MLP module.
We adopted a representation approach inspired by TabularNet~\cite{du2021tabularnet} for understanding tabular data. This approach employs Bi-GRU to aggregate information from each cell, along with its corresponding row and column. 
Moreover, our framework improves TabularNet by employing a Bi-LSTM on rows and columns, enabling the aggregation of information for the entire table, as shown in Fig.~\ref{fig:model}.

We utilize a Bi-LSTM along the rows to gather feature information for each row. Subsequently, the information from each row is consolidated to derive the overall information for the entire table. Likewise, a similar procedure is conducted along each column to obtain comprehensive table information. This aggregated information serves as input for the subsequent MLP, facilitating our model's ability to handle variations in the number of data cells.

As the positions of the tabular data items change with the table transformations in the first stage, it becomes imperative to update their positions accordingly. 
To guarantee that the data items are arranged by their original numbering, each input tabular data is assigned a unique identifier. This arrangement ensures that even if the position of the data changes, the items can still be organized based on their original numbering and then sent to the MLP module to derive the feature vector $h_D$ for the data. 

Consequently, the feature vector $h_H$ is obtained for the table headings, and the feature vector $h_D$ is obtained for the table contents. After concatenating them along the last dimension, the feature vector $h_T = \operatorname{cat}(h_H, h_D)$ is obtained for the entire hierarchical table.

\subsubsection{Agent Network Architecture}
After obtaining the representation of hierarchical tables, a series of actions related to table transformation and visualization region selection are performed on the table. 
As depicted in the agent module presented in Fig.~\ref{fig:model}, the features extracted from hierarchical tables are concatenated and fed into an Actor and a Critic network. 
The Actor network generates actions that manipulate the hierarchical table, represented by $a_1$ and $a_2$. 
The Critic network evaluates the system's current state and assesses its quality, represented as $v$. This evaluation is utilized for loss computation and model training.
The Actor-Critic architecture serves as a classic approach that has formed the basis for the development of numerous subsequent on-policy reinforcement learning algorithms.

The actions generated by the Actor are then applied to the environment, where they lead to modifications in the table.
Visualizing hierarchical tabular data encompasses a substantial action space, as explained in Sec.~\ref{sec:action-space}, and the visualization process relies on the results of table transformations. 
Therefore, we divide the training process into two stages, the table transformations in the first stage and the visualization embedding in the second stage, as shown in Fig.~\ref{fig:model}. 
The entire episode length is fixed. We define the stage ratio (\textit{SR}) as the ratio between the steps of two stages to control the number of steps within each stage.
Specifically, the actions---\textit{transpose}, \textit{aggregate}, \textit{stack}, \textit{unstack}, \textit{swap}---are actions associated with the transformation stage, and they are masked during the second stage. 
Similarly, \{\textit{row}$_{\textrm{up}}$, \textit{row}$_{\textrm{down}}$, \textit{row}$_{\textrm{left}}$, \textit{row}$_{\textrm{right}}$, \textit{col}$_{\textrm{up}}$,  \textit{col}$_{\textrm{down}}$, \textit{col}$_{\textrm{left}}$, \textit{col}$_{\textrm{right}}\}$ are regarded as actions in the selection stage, which are masked during the first stage.
Consequently, the Actor network with two action heads outputs actions for the two stages, as $a_1$ and $a_2$, representing table transformation and visual region selection operations, respectively.

\subsubsection{Auxiliary Reward}

Based on the aforementioned definition of hierarchical table visualization and the reward design, the two stages, which encompass transformation and visualization, employ different reward mechanisms.
In the first stage, tabular data transformation lacks a reward system, making it challenging to evaluate the quality of the resulting transformations.
Furthermore, even in the visualization stage, rewards are sparse because they are only acquired when insights are discovered, with no rewards if no insights are found.
This poses a challenge for the agent in determining which actions to take to potentially obtain rewards that may occur several steps later. Numerous studies have leveraged reinforcement learning techniques for visualization authoring~\cite{tang2020plotthread, deng2022dashbot, bar2020automatically}. However, these studies often neglect the issue of sparse rewards in reinforcement learning, wherein the agent receives rewards infrequently from the environment.

We introduce an auxiliary (intrinsic) reward mechanism to assist the agent in exploring new and unknown states in the environment, including new heading structures and visualization states. To accomplish this, we have employed the random network distillation technique~\cite{burda2019exploration}, which predicts the errors in the environmental state and steers the agent's exploration.  The state of the hierarchical tables encompasses both the heading structure and the visualization state, as explained in Sec.~\ref{sec:feature_extraction}, where GCN and Bi-LSTM networks were utilized. In this context, two distinct networks are deployed to predict the modifications in the hierarchical structure and the visualization state.

As illustrated in Fig.~\ref{fig:model}, both GCN and Bi-LSTM of the Curiosity Module consist of two identical networks with distinct parameters, generating different feature representations for the same environmental state. 
For representing table headings, we employ two networks: 
$f_H: H \rightarrow R^k$
and $\hat{f}_H: H \rightarrow R^k$, where the target network $f_H$ is utilized to provide a fixed representation for headings, while the predictor network $\hat{f}_H$ is trained to learn the embedding of headings.
Similarly, $f_D$ and $\hat{f}_D$ represent the target network and predictor network, respectively, for the corresponding table content.
Upon obtaining the feature vectors embedded by the four neural networks, the intrinsic reward is computed as the mean squared error between these feature vectors.
Specifically, the intrinsic reward for the table heading $r_H^{int}$ and for data items $r_D^{int}$ in the table content are defined as follows:
$$
\begin{gathered} 
r_H^{int}=\|\hat{f}_H(x)-f_H(x)\|^2\\
r_D^{int}=\|\hat{f}_D(x)-f_D(x)\|^2
\end{gathered}
$$
where $\hat{f}_H(x)$ and $\hat{f}_D(x)$ are the feature vectors obtained by the predictor networks for headings and content, $f_H(x)$ and $f_D(x)$ are the feature vectors obtained by the target networks for headings and content.

The target networks contain parameters that are randomly initialized and remain fixed during training, whereas the predictor networks have parameters that are randomly initialized and get updated during training.
Throughout the training process, the predictor networks encounter more unexplored states, and their state representations gradually converge toward the fixed representation of observations provided by the target networks. 
Consequently, the intrinsic reward computed between these two representations diminishes. 
In other words, when a state is visited less frequently, the difference between the representations of this state by the predictor networks and target networks is pronounced, thereby yielding a larger intrinsic reward to promote explorations.

In addition, we employ a joint optimization technique Extrinsic-Intrinsic Policy Optimization~\cite{chenredeeming} that combines auxiliary rewards to mitigate the necessity for manual adjustments to the relative importance of intrinsic and extrinsic rewards during the training process.
In some cases, incorporating intrinsic rewards can stimulate exploration, guiding the agent to discover unexplored states. However, in other cases, it might interfere with the primary optimization objectives. Optimizing mixed objectives does not necessarily produce the optimal policy solely based on extrinsic rewards.
The following outlines the optimization objective:
\begin{align*}
\max_{\pi_{E+I} \in \Pi} J_{E+I}\left(\pi_{E+I}\right) 
\quad & \text{s.t.}, \ J_E\left(\pi_{E+I}\right) -\max _{\pi_E} J_E\left(\pi_E\right)=0 \\
\text{where} \quad J_{E+I}(\pi_{E+I}) &=\mathbb{E}_{\pi_{E+I}}\left[\sum_{t=0}^{\infty} \gamma^t\left(r_t^E+\lambda r_t^I\right)\right], \\
J_E(\pi_{E+I}) =\mathbb{E}_{\pi_{E+I}}&\left[\sum_{t=0}^{\infty} \gamma^t r_t^E\right], \ 
J_E(\pi) =\mathbb{E}_\pi\left[\sum_{t=0}^{\infty} \gamma^t r_t^E\right]
\end{align*}
where $E$ and $I$ denote extrinsic and intrinsic, $\pi$ represents a policy to output action, $\pi_{E+I}$ represents a mixed strategy policy, $\pi_E$ represents an extrinsic strategy policy, $J_{E+I}$ represents a mixed optimization objective, $J_E$ represents an extrinsic optimization objective, $\gamma$ represents a discount rate (set to $0.99$), $t$ represents the time step, and $r_t^E$ and $r_t^I$ denote extrinsic and intrinsic rewards.

The objective is to optimize a mixed policy based on the joint optimization approach that consists of both intrinsic and extrinsic rewards. 
To ensure the extrinsic optimality, an extrinsic optimality constraint is imposed, which demands that the mixed policy outperforms the extrinsic policy in terms of the extrinsic reward objective.
In other words, the mixed policy optimized with mixed rewards must be at least as effective as a regular extrinsic policy.
As a result, the loss computation and model training are conducted based on the model's output of $r^{int}$, $r^{ext}$, $v$, $a_1$, and $a_2$.

\section{The InsigHTable System}

We have designed and developed the InsigHTable system based on the existing HiTailor system to support users to create hierarchical table visualizations with abundant data insights effectively.
More specifically, we borrow the hierarchical table visualization panel (Fig.~\ref{fig:case2}(A)) and the visualization configuration panel (Fig.~\ref{fig:case2}(D)) from HiTailor, and further complement the insight list panel (Fig.~\ref{fig:case2}(B)) and the alternative insight panel (Fig.~\ref{fig:case2}(C)).
In addition, we also design different visual cues in the hierarchical table visualization panel to facilitate users understanding the extracted data insights.

\subsection{Design Considerations}
We identified the following three design considerations for the InsigHTable system. These considerations are derived from existing studies about visualization authoring~\cite{deng2022dashbot, wu2020mobilevisfixer} and visual data storytelling~\cite{shi2020calliope, zhou2021table2charts} driven by DRL, as well as the mixed-initiative visualization systems~\cite{horvitz1999principles, wongsuphasawat2015voyager, wongsuphasawat2017voyager}.

\textbf{DC1: Enable interactive steering to guide the construction of hierarchical table visualizations.}
While the intelligent agent of the DRL algorithm autonomously generates visualization results, it may not always meet the needs of all users~\cite{deng2022dashbot, shi2020calliope, wu2020mobilevisfixer}. 
Moreover, users' interests may evolve as they explore the data.
Hence, it is essential to allow interactive steering to influence the visualization generation process~\cite{wongsuphasawat2015voyager, wongsuphasawat2017voyager}.
More specifically, the system should provide flexible interactions to allow users to remove, add, and update insights based on the generated hierarchical table visualizations~\cite{horvitz1999principles}.

\textbf{DC2: Clearly and effectively present multiple insights extracted from data in the visualization results.} 
The data visualizations are collaboratively crafted with input from both the intelligent agent and data analysts rather than being created manually.
Therefore, it is crucial for the system to ensure that the generated visualizations clearly convey the extracted data insights and avoid ambiguity~\cite{shi2020calliope, zhou2021table2charts}. 
Additionally, since hierarchical table visualizations embed many visual representations into original tables, navigating through multiple visualizations can be cognitively demanding~\cite{wongsuphasawat2015voyager, wongsuphasawat2017voyager}.
Hence, the system should strive to keep visualizations as simple as possible to prevent visual clutter and reduce the cognitive load of exploring multiple insights on a single screen.

\textbf{DC3: Maintain working memory for constructing hierarchical table visualizations.}
Systems should retain a memory of recent interactions and offer mechanisms that enable users to locate specific visual elements efficiently~\cite{horvitz1999principles}. 
This construction process of hierarchical table visualizations can be broken down into multiple steps, with each step representing a distinct state. 
Therefore, it is essential to illustrate the intermediate states.

\subsection{User Interface and Interaction}
Driven by the above design considerations, we design and develop the InsigHTable system to support users in exploring the hierarchical tabular data interactively.

InsigHTable presents the raw hierarchical tables uploaded by users within the hierarchical table visualization panel at the beginning.
After users click the ``insight recommendation'' button \raisebox{-2pt}{\includegraphics[height=1em]{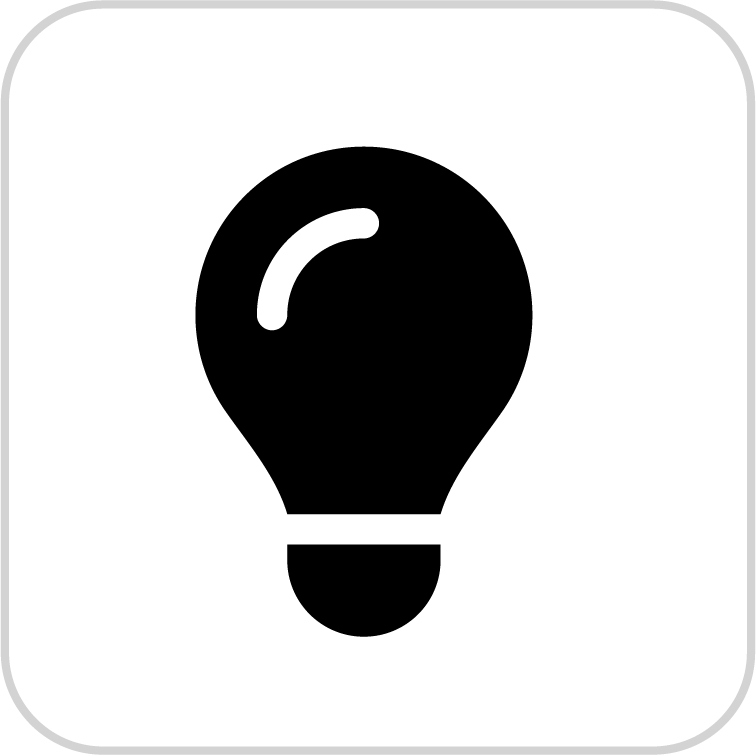}}, the DRL model generates hierarchical table visualizations with abundant data insights.  
InsigHTable employs a mixed-initiative approach based on DRL to construct hierarchical table visualizations. 
The construction process is displayed in the insight list panel (\textbf{DC3}).
This approach involves two user interactions designed to enhance and refine the generated results (\textbf{DC1}). 
First, once users find certain generated data insights unsatisfactory, they are able to remove them from the insight list panel and replace them with manually created insights. 
In the insight list panel, the generated data insights generated by the intelligent agent \raisebox{-2pt}{\includegraphics[height=1em]{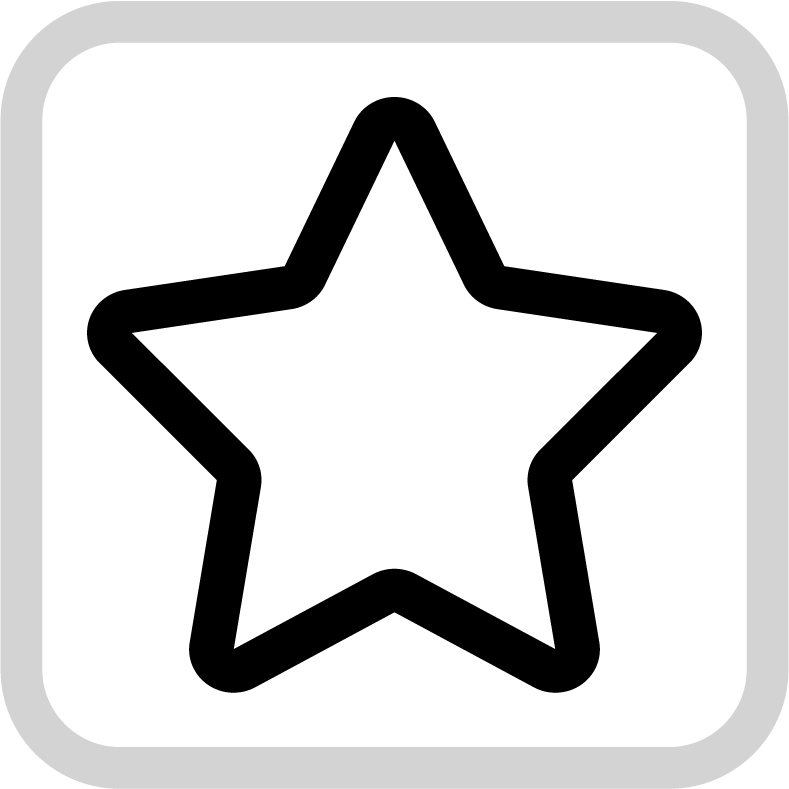}} and constructed manually \raisebox{-2pt}{\includegraphics[height=1em]{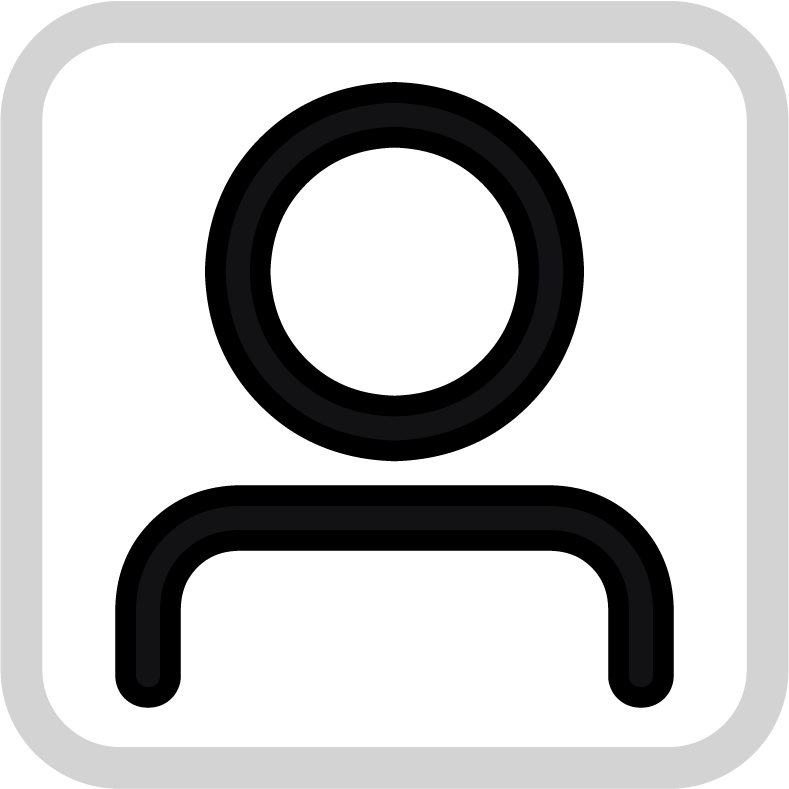}} are encoded with different icons (\textbf{DC3}). 
Second, InsigHTable can provide multiple types of data insights related to the same data cells. 
When users click on one data insight, other insights extracted from the same data block are presented in the alternative insight panel. 
Users can then choose insights from the panel and replace them with the original data insight in the hierarchical table visualization.
Subsequently, the DRL framework updates the generated data insights accordingly. 
Through these iterative processes, users can get a satisfying hierarchical table visualization. 
In particular, considering the limited space, the system does not provide the axes and legend for the embedded visual representations to avoid visual clutter and complements this information in the alternative insight panel (\textbf{DC2}).

\begin{figure*}[!h]
    \centering
    \includegraphics[width=\textwidth]{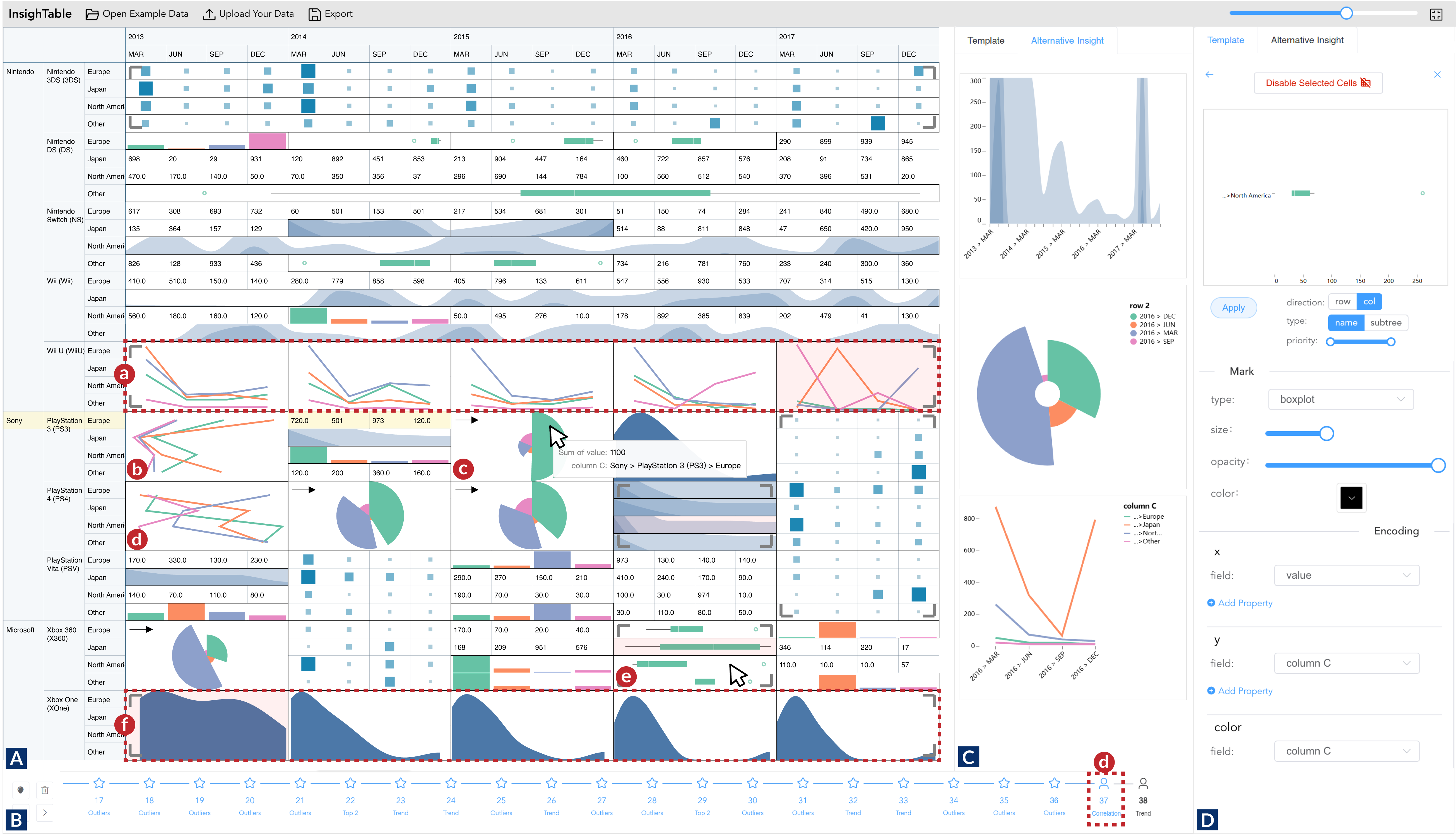}
    \caption{The hierarchical table visualization is based on a dataset of console sales and is created using the InsigHTable prototype system. This data is about sales data for Nintendo, Sony, and Microsoft, and encompasses generated visualizations that are embedded in the table content. The InsigHTable prototype system consists of four panels: (A) table visualization panel; (B) insight list panel; (C) alternative insight panel; and (D) visualization configuration panel.}
    \label{fig:case2}
\end{figure*}

The constructed visualizations consist of both single-block and multiple-block insights. 
Different from single-block insights, multiple-block insights are related to data items obtained through recommendations. 
InsigHTable incorporates various visual cues to indicate related data blocks  (\textbf{DC2}). 
For the name-based recommendation mechanism, the system recommends multiple data blocks with the same label in the heading. 
Because these blocks may not be adjacent to each other, a light-gray background is added to mark them. 
Conversely, data blocks determined by the topology-based mechanism are adjacent. 
The system adds rectangular corner markers \raisebox{-2pt}{\includegraphics[height=1em]{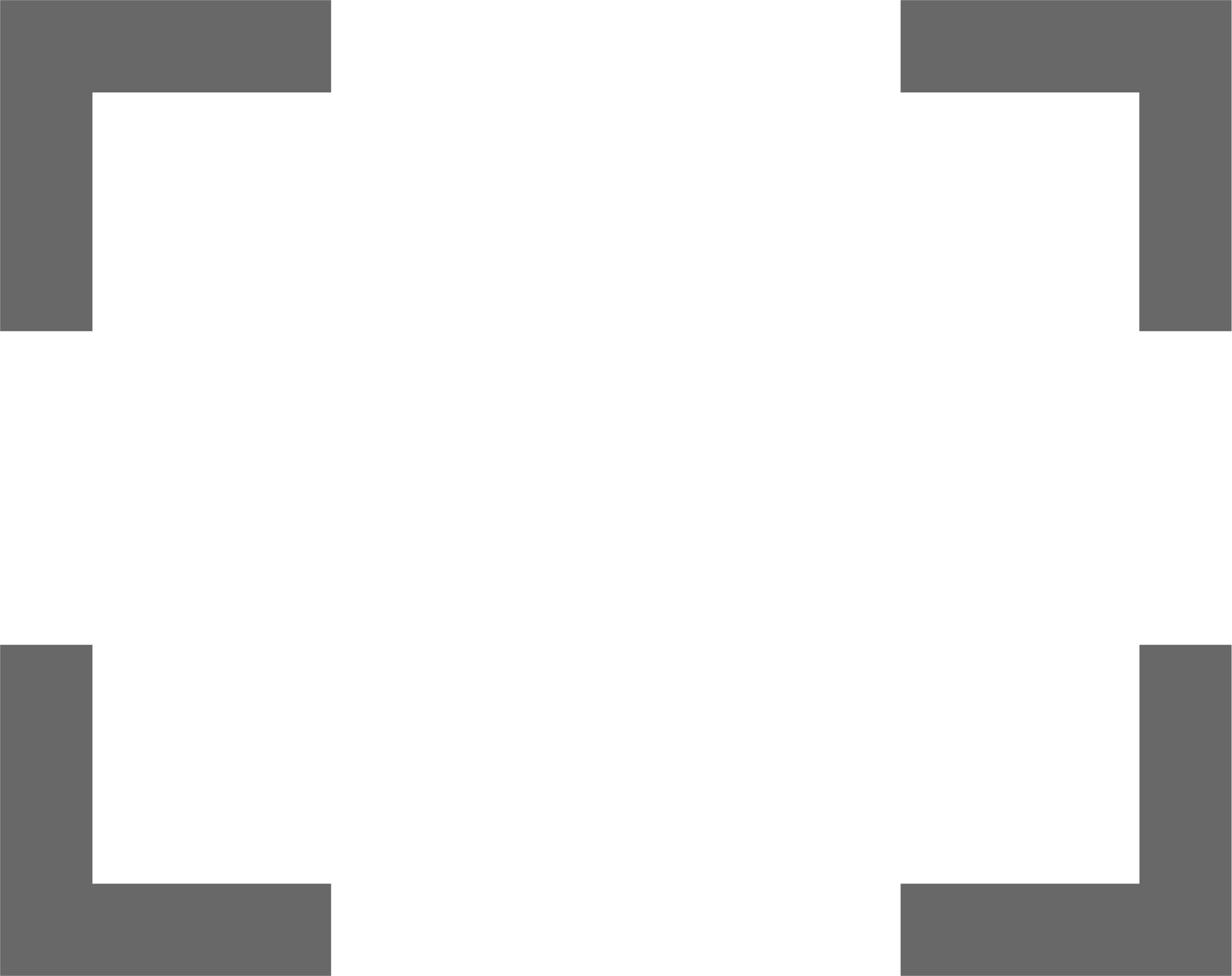}} to denote these related blocks.
Based on the insight explanations provided in Sec.~\ref{sec:insight}, multiple-block insights are derived from insights detected based on a single block and whether insights in the same category exist in all recommended data blocks. Therefore, to indicate data blocks that have a distinct insight compared to others, a light-red background is utilized. 
From the insight list panel, users can quickly locate the insight of interest by clicking on the icon.
For each visualization embedded in the table content, the data analyst can check the underlying data attribute by hovering over the visual elements. 
For example, when hovering over a visual element, the corresponding row is highlighted in yellow, as demonstrated in Fig.~\ref{fig:case2}(c).

\section{EVALUATION}
To evaluate the effectiveness of InsigHTable, we conduct two case studies and quantitative experiments.
The quantitative experiments encompass hyperparameter tuning, ablation studies, and comparisons with heuristic methods.

We train the reinforcement learning model based on the HiTab~\cite{cheng2022hitab} dataset, which is utilized for question answering and natural language generation over hierarchical tables.
The parameters for the training reinforcement learning model are as follows: the learning rate is $2 \times e^{-4}$, the number of parallel environments is $128$, the interval for updating parameters is $64$ steps, the mini-batch size is $512$, the number of total time steps is 500M, the clip coefficient for PPO is $0.2$, and the value coefficient is $0.5$. The stopping condition for the DRL framework is when the episode exceeds the threshold, which we set to $200$ in our experiment, or when all cells within tabular data are embedded with visualizations. 

\subsection{Case Studies}
The case studies illustrate two real-world examples of insight extraction and visualization using InsigHTable, one console sales dataset, and one insurance statistical dataset.

\subsubsection{Case Study 1: Console Sales Data Analysis}
In the first case study, we collaborated with a data analyst to analyze market sales data.
The constructed visualizations are depicted in Fig.~\ref{fig:case2}. 
The underlying data of this result is about the sales of console, including sales data for gaming consoles from different brands, categories, and regions. 
The data also consists of sales across various regions and time periods.

To begin the exploration process, the data analyst clicked the ``insight recommendation'' button \raisebox{-2pt}{\includegraphics[height=1em]{figures/lightbulb.png}} in the insight list panel (Fig.~\ref{fig:case2}(B)), and the visualizations corresponding to the insights are then directly embedded into the corresponding cells/blocks within the table content. 
Additionally, the data analyst clicked the ``\textit{previous}'' \raisebox{-2pt}{\includegraphics[height=1em]{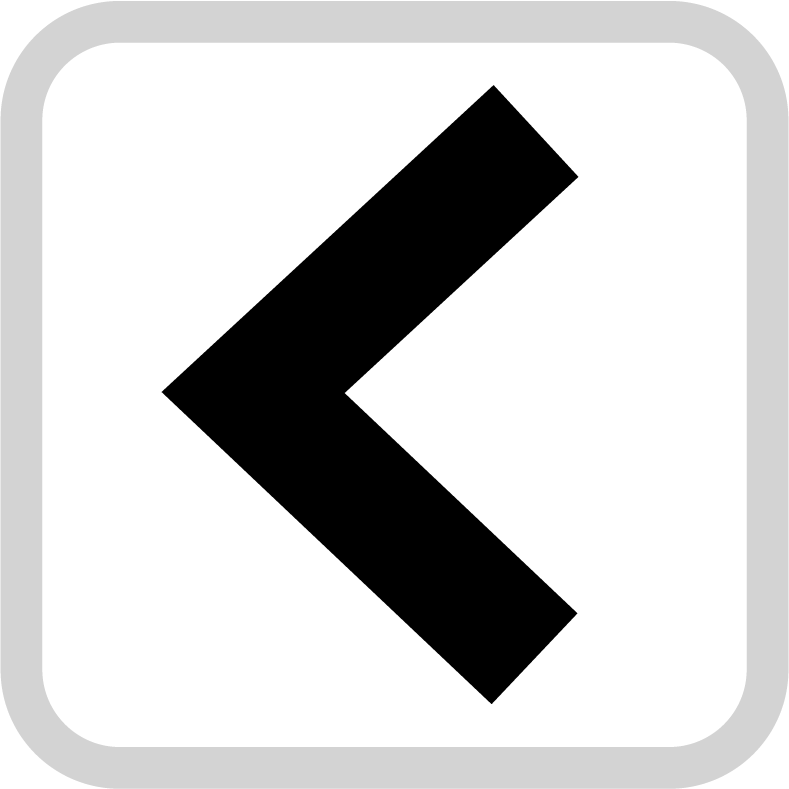}} and ``\textit{next}'' \raisebox{-2pt}{\includegraphics[height=1em]{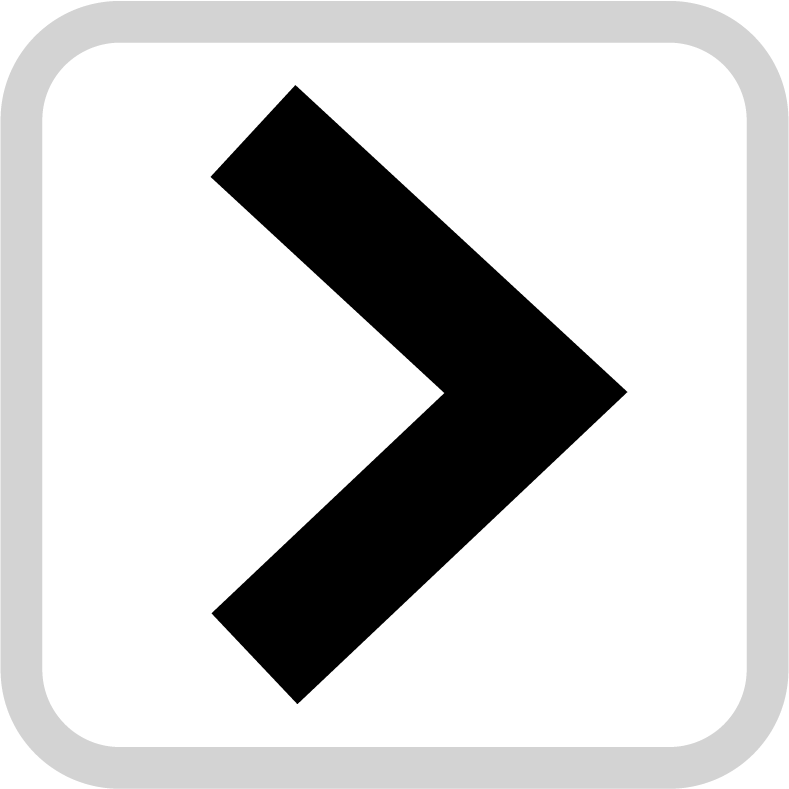}} buttons located on the left of the insight list to update the insights displayed within the hierarchical table visualization panel.

The visualization results constructed by the data analyst are shown in Fig.~\ref{fig:case2}. 
These visualizations reveal multiple intriguing data insights. 
First, the visualization in Fig.~\ref{fig:case2}(c) demonstrates the proportion of PlayStation 3's sales in 2015, and the sales in Europe are dominant among all regions, which is different from the sales of PlayStation 4.
Fig.~\ref{fig:case2}(e) indicates that only Japan did not have an outlier for the sales of Xbox 360 in 2016.
The data analyst further checked its alternative insights, including change point insight (\textit{shape insight}), outstanding top two insight (\textit{point insight}), and correlation insight (\textit{compound insight}). It found that the sales' trend in Japan are significantly different from the other regions. 

As shown in Fig.~\ref{fig:case2}(a), the sales trends in 2017 are different from those in other years. 
Notably, a sharp rise in sales of the Wii U console in Japan during this period. 
Upon investigation, the data analyst discovered that this surge was attributed to the cessation of Wii U console production at that time, leading to a substantial increase in demand for memory in Japan. 
Furthermore, as illustrated in Fig.~\ref{fig:case2}(f), the sales distribution in 2013 deviates notably from the patterns observed in other years. 
After analyzing the underlying data, the data analyst found that this discrepancy was primarily because of the distinct sales performance of the Xbox One in America during that year. 
The analysts' discovery prompted them to determine that the initial response to the Xbox One in America was fairly unenthusiastic.
However, after establishing a gaming console ecosystem, the sales of Xbox One exhibited significant growth.

Based on the constructed hierarchical table visualizations, the data analyst also created data visualizations manually, which is shown at the end of the insight list panel. 
The visualization in Fig.~\ref{fig:case2}(d) is to explore the difference between PlayStation 3 and PlayStation 4. 
Fig.~\ref{fig:case2}(b) indicates that the sales of PlayStation 3 among different regions have correlation data insights. 
However, after creating the visualization in Fig.~\ref{fig:case2}(d), the data analysts found that the correlation insight does not exist in PlayStation 4. 
Such a difference inspires the data analysts to explore the different strategies of these two products in different regions.

\subsubsection{Case Study 2: Insurance Statistics Data Analysis}

\begin{figure*}[ht]
    \centering
    \includegraphics[width=\textwidth]{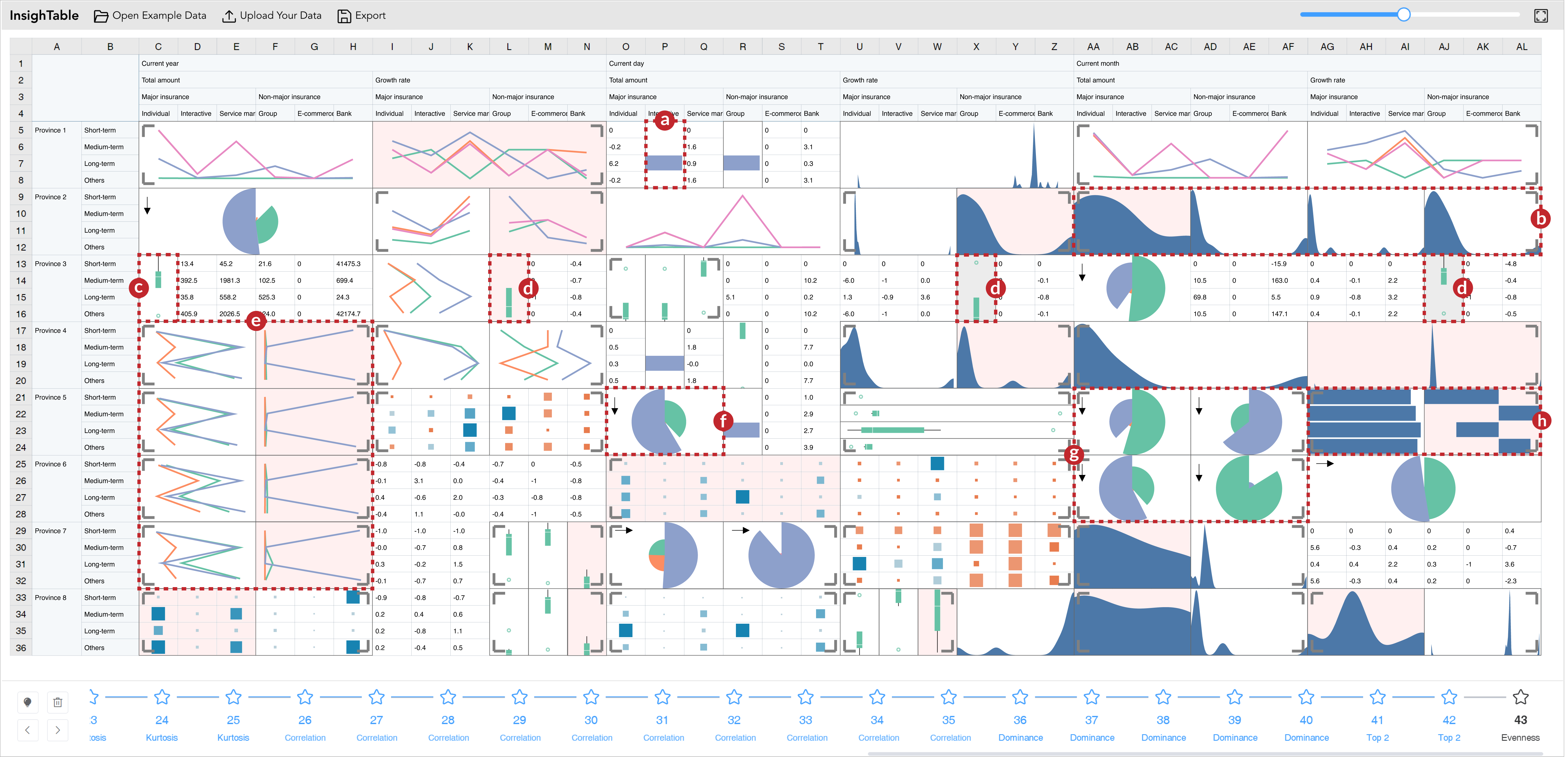}
    \caption{One hierarchical table visualization generated by InsigHTable: the premium statistical data of an insurance company with various visualizations corresponding to different insights embedded in the hierarchical tabular data. Several visual cues indicate the related blocks: the rectangular corner markers indicate whether the data blocks are recommended using the topology-based mechanism, and the light-gray background indicates that the data blocks are recommended using the name-based mechanism. The light-red background indicates a distinct insight exists within the block. The bottom of the figure shows the generated insight list.}
    \label{fig:case}
\end{figure*}

We conducted the second case study with data analysts in a well-known insurance company that heavily relies on hierarchical tabular data. 
The findings demonstrate that InsigHTable can significantly facilitate the extraction of valuable insights from these tables, enabling data analysts to work efficiently and resulting in rich data insights.

The analysts chose one hierarchical tabular dataset related to premium statistics, which contains information on premium income and growth rate for each province. The attributes in the headings of this hierarchical table include province, premium type, classification of sales channel, time range, and type of statistic value (total amount or growth rate).
Fig.~\ref{fig:case} shows an example of hierarchical table visualization, which demonstrates the extracted insights by InsigHTable.

For single-block insights, as shown in Fig.~\ref{fig:case}(a), the total amount of the long-term insurance in $p1$ (province 1) is an outlier within the single column, which exceeds the power-law distribution of the column. This insight is visualized using a bar chart visualization. 
As shown in Fig.~\ref{fig:case}(c), one data item is much smaller (less than $Q1-3\times IQR$) than the others within the single column. After checking the underlying data, the analyst finds that the total amount of the medium-term insurance in $p3$ is an outlier.
Fig.~\ref{fig:case}(f) presents a data block with a dominance insight. Within the data block, the data items are aggregated by computing the average value along the column (indicated by the downward-pointing arrow \raisebox{-2pt}{\includegraphics[height=1em]{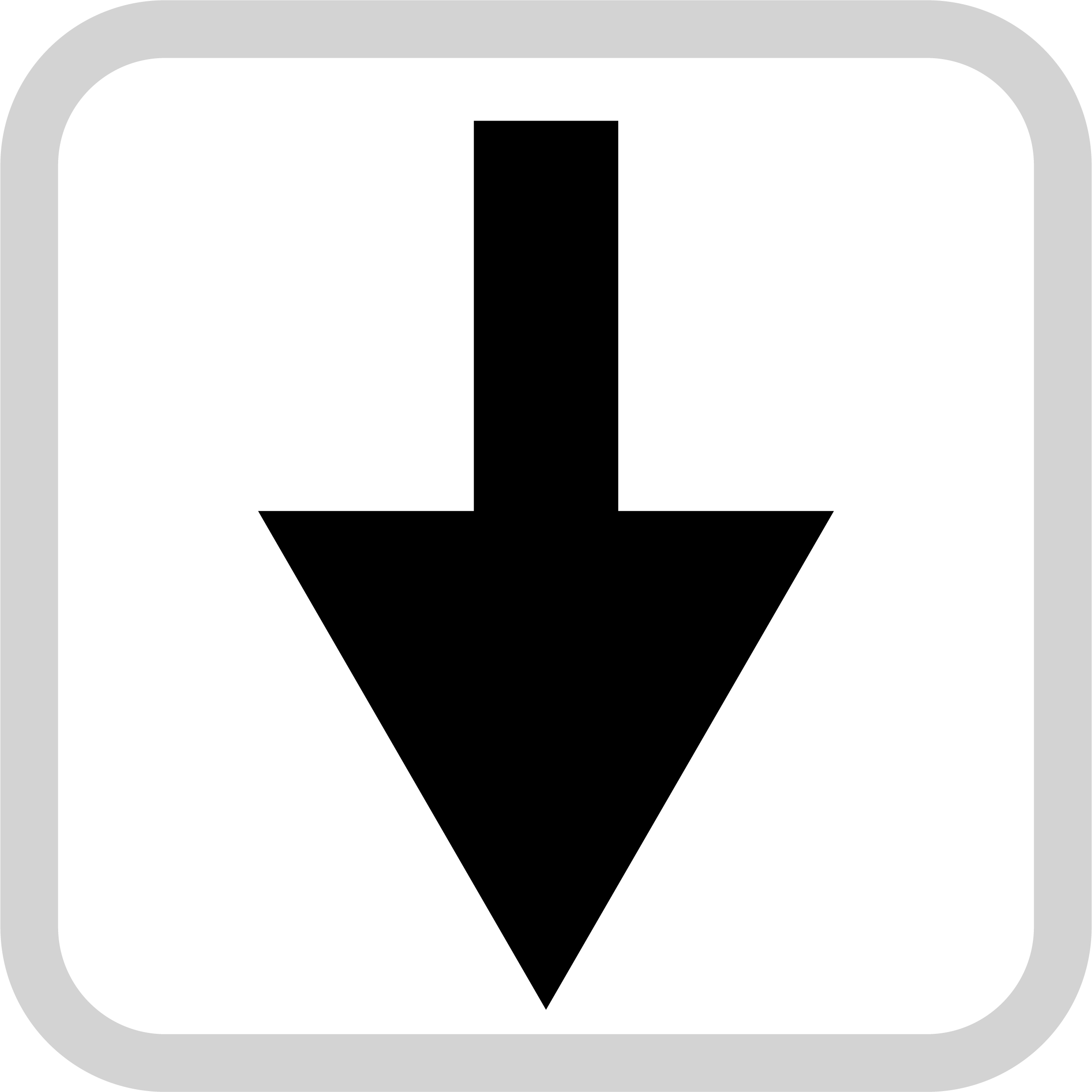}}). This insight is visualized using a radial plot visualization. The analyst finds that the service marketing channel category represents the largest proportion of channels.

The visualization results comprise various insights across multiple blocks.
Fig.~\ref{fig:case}(d) demonstrates three recommended data blocks through the name-based mechanism.
These data blocks share identical labels except for the top level in the column headings, which represent the current year, month, and day. The analyst discovered that both data blocks corresponding to the current day and month contain an outlier, whereas the data block corresponding to the current year does not. This suggested that the growth rates of the four different insurance types have displayed increasing variability in recent times.

In addition to that, Fig.~\ref{fig:case}(b) depicts the skewness insight related to four recommended data blocks based on the topology-based mechanism. 
Among these four data blocks, the analyst found that the first block has a data distribution that is not significantly skewed, while the remaining three blocks exhibit strong skewness. This suggested that the data distribution is large in scale but is highly concentrated within a narrow range.
This information informed the analyst that the total amount of the sale channels under the major insurance category is more balanced, whereas the other channels need to be improved. 

Furthermore, Fig.~\ref{fig:case}(e) shows the data blocks determined by the topology-based mechanism. The multi-series line chart on the left block illustrates the correlation between the sum of the insurance sale channels under the major insurance category, whereas, on the right, no such correlation exists. This insight encouraged the analyst to analyze the underlying relationships between these sale channels and improve the strategies, for example, avoiding the interrelated effects between different types of insurance.

Moreover, upon aggregating the data items within multiple blocks along the row in Fig.~\ref{fig:case}(h), the analyst identifies the evenness insight. 
The left block exhibited the evenness characteristic, while the chart on the right does not, implying that the growth rate of the ``major insurance'' in $p5$ is similar, whereas the difference in ``non-major insurance'' is significant at the current month. 
Therefore, it is necessary to specify policies to improve the efficiency of non-major insurance.

Similarly, Fig.~\ref{fig:case}(g) presents four data blocks, each comprising an aggregation insight. The top two data blocks are a set of recommended insights, while the bottom two data blocks are another set.
All four insights belong to the top two or dominance category and are visualized using radial plot visualization. 
The analyst found that distributions of major and non-major insurance categories vary significantly between p5 and p6. 
Specifically, under the major insurance category, short-term insurance and medium-term insurance of p5 and p6 have the largest and second-largest values, while the relative sizes are different. Under the non-major insurance category, the dominance category is short-term insurance in p5, whereas the dominance category is the medium category in p6.
This data insight suggested the analysts analyze the variance of strategies in detail between $p5$ and $p6$ to realize complementary.

\subsection{Quantitative Experiments}
We perform four experiments to validate the effectiveness of the InsigHTable system, including a hyperparameter tuning experiment, an ablation study, a robustness evaluation, and a comparison with heuristic insight extraction methods to show the superiority and necessity of the DRL framework.
All experiments are conducted using the three diversity metrics (IR, AR, and ER) defined in Sec.~\ref{sec:reward-function}. 
The experiments are conducted on a machine equipped with an Intel(R) Xeon(R) Platinum 8280 CPU@2.70GHz featuring 112 cores, 8 NVIDIA RTX A6000 GPUs, and 754GB of memory.

\subsubsection{Hyperparameter Tuning}

In the context of DRL, the effectiveness of neural network models heavily relies on hyperparameters, which are crucial settings influencing the model's performance. Properly tuning these hyperparameters is essential to achieve optimal results and enhance the overall learning process.

The ratio between the number of executed steps in the transformation and the whole visualization stages denoted as \textit{SR} is a crucial hyperparameter in the DRL framework. 
In our experiment, we carefully adjusted these hyperparameters and conducted cross-validation experiments to ensure optimal performance. The results of these experiments are presented in Table~\ref{table:hyperparameter}. 
Adjusting the value of \textit{SR} is to strike a balance between the two stages. A small value limits the number of transformation steps, resulting in an ineffective transformation of the hierarchical table and notably inferior performance. Conversely, a large value increases the number of transformation actions, resulting in fewer data selection and visualization embedding actions by the agent and diminished performance. The experiment results show that the model performances are optimal when the \textit{SR} value is assigned as $0.04$.

The number of GCN layers denoted as \textit{L} also affects the performance of the model. More specifically, a small number of GCN layers might diminish the model's ability to represent the tabular data's hierarchical structure.
Conversely, many GCN layers lead to an increase in network parameters, which makes the learning of the model challenging, resulting in poor performance. Therefore, choosing the appropriate number of layers is crucial to ensure optimal performance. The experiment demonstrates that a moderate number of three layers is appropriate. Note that reinforcement learning is different from supervised learning in that an excessive number of network parameters can lead to great difficulties for model training.

\begin{center}
\begin{table}[]
\caption{\textbf{Impact of two hyperparameters.} The IR, AR, and ER indicate Insight Ratio, Area Ratio, and Evenness Ratio, respectively. The italics in the headers and the content indicate the optimal parameters and results.}
\label{table:hyperparameter}
\centering
\begin{tabular}{cc|ccccc}
\Xhline{1.2pt}
\multicolumn{2}{c|}{{GCN Layer (L)}} & L = 1 & L = 2 & \underline{\textit{\textbf{L = 3}}}        & L = 4 & L = 5 \\ \hline
\multicolumn{1}{c|}{{Stage Ratio}} & IR & 0.2   & 0.2   & 0.2            & 0.2   & 0.2   \\ \cline{2-7} 
\multicolumn{1}{c|}{= 0.01}                           & AR & 0.425 & 0.403 & 0.337          & 0.290 & 0.031 \\ \cline{2-7} 
\multicolumn{1}{c|}{}                           & ER & 0.276 & 0.301 & 0.301          & 0.299 & 0.276 \\ \hline
\multicolumn{1}{c|}{{\underline{\textit{\textbf{Stage Ratio}}}}} & IR & 0.6   & 0.6   & \underline{\textit{\textbf{0.7}}}   & 0.6   & 0.6   \\ \cline{2-7} 
\multicolumn{1}{c|}{{\underline{\textit{\textbf{= 0.04}}}}}                           & AR & 0.637 & 0.604 & \underline{\textit{\textbf{0.714}}} & 0.674 & 0.628 \\ \cline{2-7} 
\multicolumn{1}{c|}{}                           & ER & 0.674 & 0.560 & \underline{\textit{\textbf{0.691}}} & 0.618 & 0.573 \\ \hline
\multicolumn{1}{c|}{{Stage Ratio}} & IR & 0.4   & 0.5   & 0.5            & 0.4   & 0.2   \\ \cline{2-7} 
\multicolumn{1}{c|}{= 0.08}                           & AR & 0.504 & 0.664 & 0.694          & 0.361 & 0.389 \\ \cline{2-7} 
\multicolumn{1}{c|}{}                           & ER & 0.486 & 0.558 & 0.645          & 0.587 & 0.301 \\ \hline
\multicolumn{1}{c|}{{Stage Ratio}} & IR & 0.3   & 0.5   & 0.5            & 0.4   & 0.4   \\ \cline{2-7} 
\multicolumn{1}{c|}{= 0.12}                           & AR & 0.396 & 0.365 & 0.563          & 0.301 & 0.250 \\ \cline{2-7} 
\multicolumn{1}{c|}{}                           & ER & 0.311 & 0.537 & 0.672          & 0.561 & 0.487 \\ \hline
\multicolumn{1}{c|}{{Stage Ratio}} & IR & 0.5   & 0.4   & 0.3            & 0.3   & 0.2   \\ \cline{2-7} 
\multicolumn{1}{c|}{ = 0.16}                           & AR & 0.283 & 0.053 & 0.285          & 0.419 & 0.371 \\ \cline{2-7} 
\multicolumn{1}{c|}{}                           & ER & 0.426 & 0.301 & 0.301          & 0.301 & 0.268 \\ 
\Xhline{1.2pt}
\end{tabular}
\end{table}
\end{center}

\subsubsection{Ablation Study}
We designed an ablation experiment to compare the effectiveness of the modules about GCN feature extraction, auxiliary reward methods, and two-stage action selection used in the model separately, and the results are shown in Table~\ref{table:ablation}.

The first part is about the impact of the GCN module. 
The ablation study is to remove the GCN model and does not differentiate the table headers and table content, as shown in the Agent Module and Curiosity Module in Fig.~\ref{fig:model}.
The GCN module of the DRL framework enables the structure and content of the hierarchical tables' multi-level headers are inputted into the model, which improves the effectiveness according to the experiment results. 
Without the GCN module, the model cannot properly understand heading structures to perform transformation actions.
Although the MLP module can still retrieve table contents from the input without distinguishing between headers and content, it prevents the agent from exploring various hierarchical table organizations.
Consequently, the model's performance will be worse than that of InsigHTable.

The second part is about intrinsic rewards. 
The ablation study is to remove the Curiosity Module of the DRL framework, as depicted in Fig.~\ref{fig:model}. 
The exclusion of the curiosity module and the consequent lack of intrinsic rewards have a deleterious impact on the results.
Moreover, the rewards during the hierarchical table visualization construction are sparse, with no reward being provided during the stage of hierarchical table transformation despite the existence of numerous states concerning the structure of the hierarchical tables. 
Without the intrinsic reward, the agent cannot comprehensively explore different hierarchical table organizations for embedding visual representations. Therefore, the performance of the model is also worse than that of InsigHTable.
Furthermore, when the visualization stage fails to yield any insights, the reward is also absent. The experimental findings indicate that intrinsic rewards can effectively facilitate exploration.

\begin{table}[]
\centering
\caption{\textbf{Ablation study results.} The italics in the headers and the content indicate the optimal parameters and results.}
\label{table:ablation}
\begin{tabular}{cc|ccc}
\Xhline{1.2pt}
\multicolumn{2}{c|}{} & \makebox[0.13\columnwidth][c]{Insight Ratio}  & \makebox[0.13\columnwidth][c]{Area Ratio} & \makebox[0.17\columnwidth][c]{Evenness Ratio} \\ \hline
\multicolumn{2}{c|}{\makebox[0.2\columnwidth][c]{\underline{\textit{\textbf{InsigHTable}}}}}                                               & \underline{\textit{\textbf{0.7}}} & \underline{\textit{\textbf{0.714}}} & \underline{\textit{\textbf{0.691}}} \\ \hline
\multicolumn{1}{c|}{{\makebox[0.1\columnwidth][c]{Ablation}}}  & w/o GCN  & 0.2 & 0.694 & 0.298 \\ \cline{2-5} 
\multicolumn{1}{c|}{}  & w/o $r^{int}$   & 0.2 & 0.409 & 0.271 \\ \cline{2-5} 
\multicolumn{1}{c|}{}  & w/o two stage   & 0.2 & 0.178 & 0.301 \\ 
\Xhline{1.2pt}
\end{tabular}
\end{table}

The transformation results serve as the foundation for embedding visual representations, and the transformation actions of hierarchical table may destroy the embedded visualizations. 
Consequently, it is essential to divide the construction process of tabular data visualization into two stages. 
In the ablation study, we do not differentiate the transformation stage and visualization stage, as shown in Fig.~\ref{fig:model}.  
More specifically, the transformation and visual region selection actions may be executed at any step, which could significantly impact the final tabular data visualization results.
In particular, one transformation action is designed to clear all embedded visual representations because continuous data cells within a data block may become disjointed after transformations.
Without the two-stage mechanism, the model will lead to inefficient explorations of agents.
The experimental findings reveal that the model without a two-stage mechanism yields significantly inferior results compared to other two-stage models, confirming the mechanism's effectiveness.

\subsubsection{Robustness Testing}

\begin{figure}[!ht]
    \centering
    \includegraphics[width=\linewidth]{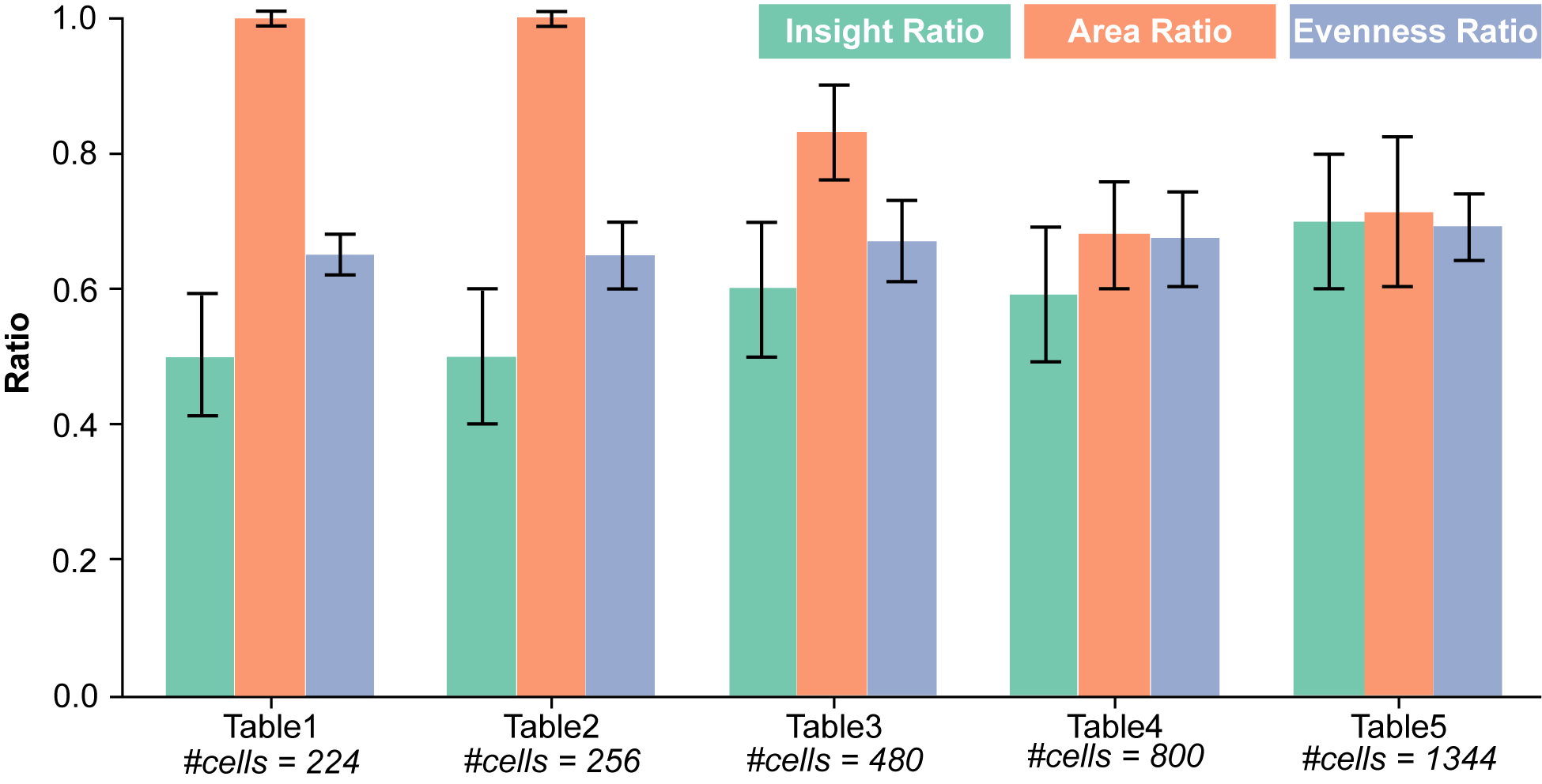}
    \caption{The bar chart indicates the Insight Ratio, Area Ratio, and Evenness Ratio results, with error bars indicating the degree of the variance with randomly initialized states of table headings.}
    \label{fig:robust}
\end{figure}

To assess the robustness of our method, we conducted an experiment using several hierarchical tables with different numbers of table cells ranging from 224 to 1344.
This experiment is to evaluate the quality of the hierarchical table visualization results with setting various heading structures of hierarchical tables. 
Specifically, we applied random transformation actions to these hierarchical tables and used the transformation results as inputs to the model.
For each hierarchical table, we selected 10 different initial states.
As illustrated in Fig.~\ref{fig:robust}, the results demonstrate that the DRL model exhibits robustness by consistently achieving satisfactory outcomes across various indicators (\textit{IR}, \textit{AR}, and \textit{ER}) even when the hierarchical table's heading structure is randomly initialized.

\subsubsection{Comparison with Baselines}

\begin{figure}[!ht]
    \centering
    \includegraphics[width=\linewidth]{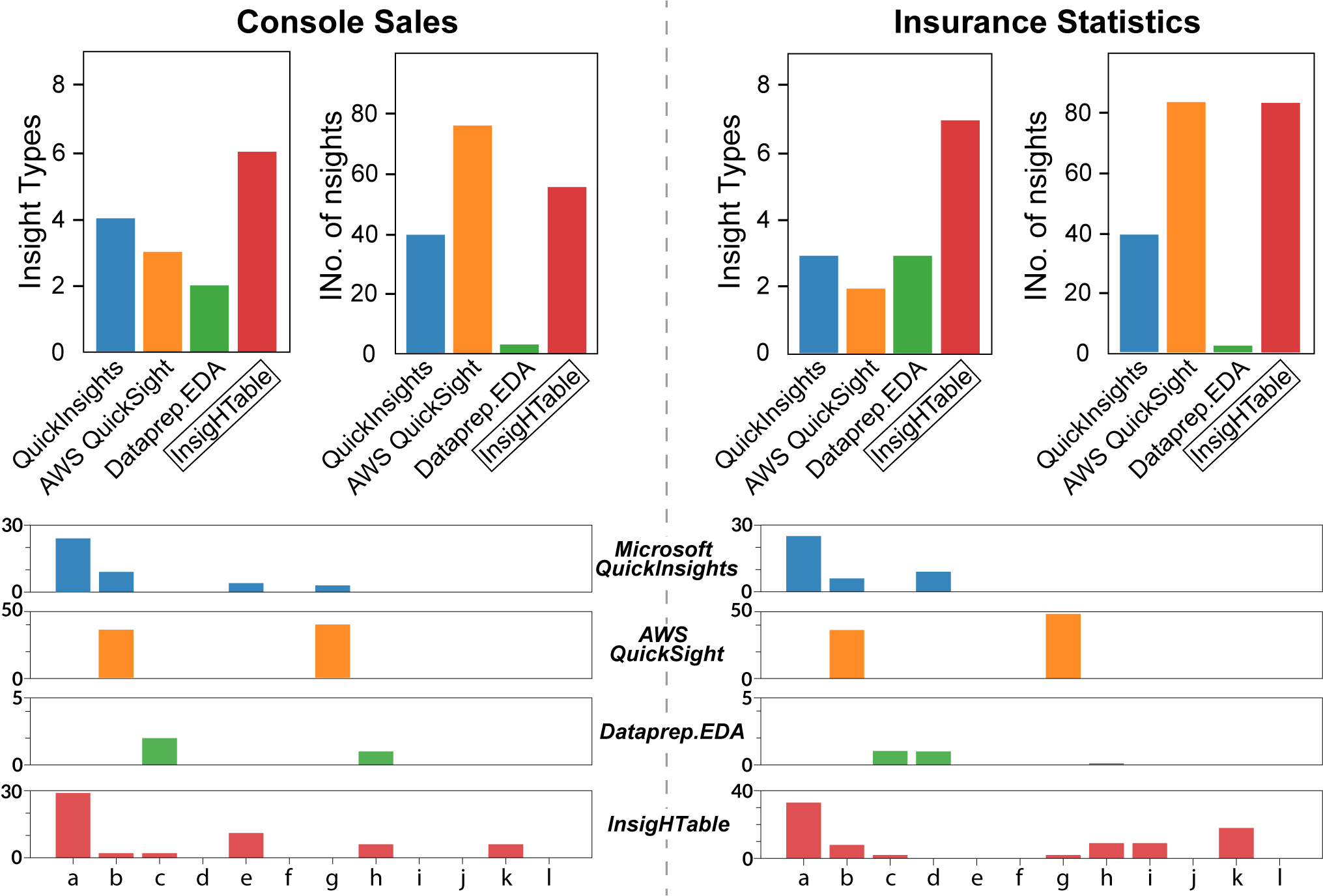}
    \caption{The experiment results are compared to three heuristic insight extraction methods, including PowerBI QuickInsights, AWS QuickSight, and DataPrep.EDA, with a metric that measures the number of insights for each insight category type. The labels $a$-$l$ indicate Outlier, Dominance, Outstanding top two, Outstanding negative, Trend, Change point, Evenness, Skewness, Kurtosis, Dependence, Correlation, and Cross-measure correlation, respectively.
    }
    \label{fig:heuristic-comparison-results}
\end{figure}

We conducted the comparative analysis with three heuristic methods employed for automatic insight extraction: PowerBI QuickInsights~\cite{ding2019quickinsights}, AWS QuickSight~\cite{quick_sight}, and DataPrep.EDA~\cite{peng2021dataprep}. 
In this experiment, we utilize hierarchical tables from real-world applications, including console sales and insurance statistics, as explained in the case studies. 
More specifically, QuickInsights on Power BI and QuickSight on AWS do not require programming; users only need to upload the tabular data and generate the insight list directly. On the contrary, Dataprep.EDA based on Python requires users to install the dataprep library and run the codes to get the generated insight list.
Based on the extracted insights, we computed the number corresponding to each insight category to evaluate the performance.

The findings demonstrate the effectiveness of the InsigHTable system in extracting data insights from hierarchical tables. 
First, it is important to note that none of the three heuristic methods are capable of the analysis of hierarchical tabular data because they are primarily designed for flat tables. 
In addition, InsigHTable allows the integration of data and visualization results by embedding visualization results into hierarchical tables, whereas the other methods generate independent visualization results. 
This highlights the superiority of our approach in the context of hierarchical tabular data scenarios.
However, as explained in Sec.~\ref{sec:introduction}, all hierarchical tables can be converted to flat tables after transformations. 
Therefore, we take these flat tables as input for the traditional insight extraction method.

The results, depicted in Fig.~\ref{fig:heuristic-comparison-results}, demonstrate the number of insights corresponding to each category. 
First, the insight types identified by the three heuristic methods are relatively limited.
Specifically, QuickInsights primarily focuses on (a)-outlier, (b)-dominance, (e)-trend and (g)-evenness insight types, and DataPrep.EDA identifies additional insight types including (h)-skewness, the percentage of zeros, and the percentage of negative numbers.
AWS QuickSight also identifies insights of only three types: (c)-outstanding top two, (d)-outstanding negative, and (h)-skewness.
However, InsigHTable is capable of extracting various insight types, as depicted in the above two use cases, including point insights, shape insights, and compound insights. Furthermore, InsigHTable enables the extraction of multiple-block insights based on the data item relationships.
Secondly, in terms of the quantity of extracted insights, AWS QuickSight is able to identify a larger number of insights, but it requires lots of users' interactive operations, for example, users need to select the interested entries from data rows and columns interactively. 
Conversely, DataPrep.EDA only identified few data insights.
Based on the experiment results, the quantity and diversity of insights extracted by InsigHTable are better than existing insight extraction methods.

Compared to heuristic methods, InsigHTable demonstrates superiority in insight extraction within the context of hierarchical tabular data transformations and visualization scenarios.

\section{DISCUSSION AND FUTURE WORK}

\textbf{Scalability.} 
The insight formulation relies on established research and the expertise of domain experts who are familiar with exploring hierarchical tables. Given the prevalence of hierarchical tables in various application scenarios, some insights related to hierarchical tables may have yet to be considered. Nevertheless, the reinforcement learning model of InsigHTable is scalable, enabling the seamless integration of new insights and visualizations (\textit{e.g.}, hierarchical data visualizations~\cite{schulz2011treevis, li2023GoTreeScape, li2020chiea}) into the system without changing the system's architecture. In the future, we plan to deploy our system online and collect more feedback to improve our insight formulation for hierarchical table data. Additionally, we aim to enhance the visualization results corresponding to the newly discovered insights. 
For the size of hierarchical tables, implementing the InsigHTable system is based on Scalable Vector Graphics, and the performance will decrease due to many DOM elements. 
We plan to improve the scalability by enabling users to filter and select subregions to narrow the exploration~\cite{lu2017interaction}.

\textbf{Interpretability.} The InsigHTable system utilizes a reinforcement learning neural network to represent hierarchical tabular data and produce actions for the transformations and visualizations. However, the reinforcement learning model is a black box for users, making it challenging to understand or explain how the agent makes these decisions. This issue is especially critical in finance application scenarios, where understanding the decision-making process can increase users' acceptance of the visualization results and enable them to identify potential errors or issues promptly.
In addition, it can also facilitate users in adjusting parameters and optimizing algorithms more effectively, thus improving the model’s performance. The lack of interpretability is a common problem for reinforcement learning, and some studies~\cite{wang2018dqnviz, metz2022comprehensive, mishra2022not} on visualization for reinforcement learning aim to explain the decision-making process. These areas are also worth exploring in our future research.

\textbf{Human-AI interaction.} The insights generated by InsigHTable are based on pre-designed patterns and do not account for users' intentions or interests. 
Although users have the ability to view and modify the results in the user interface, the reinforcement learning model may need to comprehend their intentions more effectively.
Recent advancements in human-AI interaction techniques, such as ChatGPT driven by large language models, have shown significant progress~\cite{yang2023foundation, li2024visualization}. This technique leverages the reinforcement learning from human feedback (RLHF) technique, which emphasizes the role of humans in machines and has extensive applications. In the future, we plan to explore the potential of ChatGPT or RLHF to improve the interaction between users and the reinforcement learning model and to enhance the quality of extracted insights.

\section{CONCLUSION}

In this work, we propose InsigHTable, an insight-driven method for authoring hierarchical table visualizations with rich data insights based on DRL. 
InsigHTable realizes the mixed-initiative approach for hierarchical table transformation and visual representation embedding. 
We extend the insight formulation for flat tabular data to hierarchical tabular data and define metrics to validate the quality of visualization results. 
The DRL framework of InsigHTable utilizes a two-stage approach that combines a graph neural network to represent hierarchical structure states and an auxiliary reward mechanism to encourage exploration. 
We evaluated the performance of the reinforcement learning model using two case studies and quantitative experiments. The results demonstrate the effectiveness of InsigHTable.

\section*{Acknowledgments}
We sincerely thank Prof. Michael McGuffin at ETS for his great help on this work. This work is supported by National Key R\&D Program of China (2021YFB3301500), NSFC (62302038, U2268205), Young Elite Scientists Sponsorship Program by CAST (2023QNRC001), China Railway Group Co. Ltd. Science and Technology Research and Development Program (N2022J014), and the Center of National Railway Intelligent Transportation System Engineering and Technology (RITS2022KF03).

\bibliographystyle{IEEEtran}
\bibliography{insightable}

\newpage

\section{Biography Section}

\begin{IEEEbiography}
[{\includegraphics[width=1in,height=1.25in,clip,keepaspectratio]{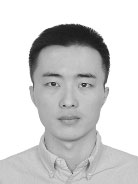}}]
{Guozheng Li}
received his PhD degree in Computer Science from the school of EECS, Peking University in 2021. He is currently an assistant professor with the School of Computer Science
and Technology, Beijing Institute of Technology, China. His major research interests include information visualization, especially hierarchical data visualization and visualization authoring.
\end{IEEEbiography}

\begin{IEEEbiography}
[{\includegraphics[width=1in,height=1.25in,clip,keepaspectratio]{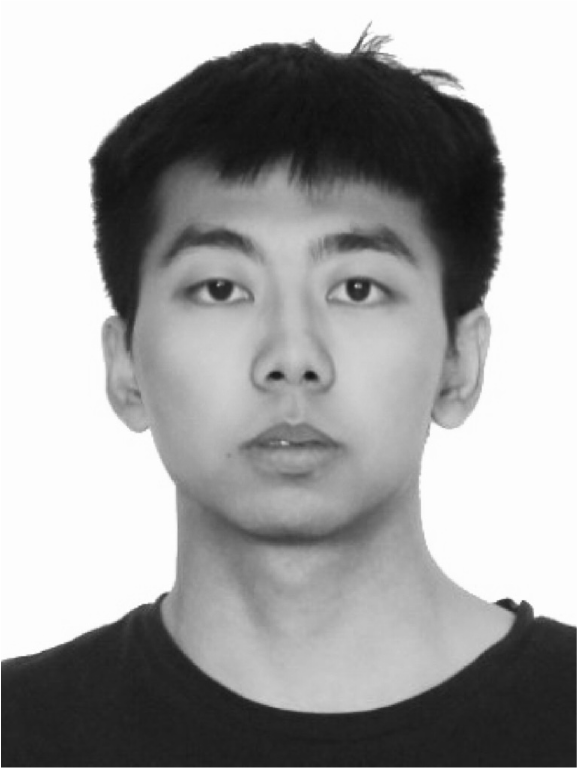}}]
{Peng He}
received his BS degree in Computer Science from Beijing Institute of Technology in 2021.
He is currently a master student at the School of Computer Science and Technology at the Beijing Institute of Technology, China.
His major research interests include reinforcement learning and information visualization.
\end{IEEEbiography}

\begin{IEEEbiography}
[{\includegraphics[width=1in,height=1.25in,clip,keepaspectratio]{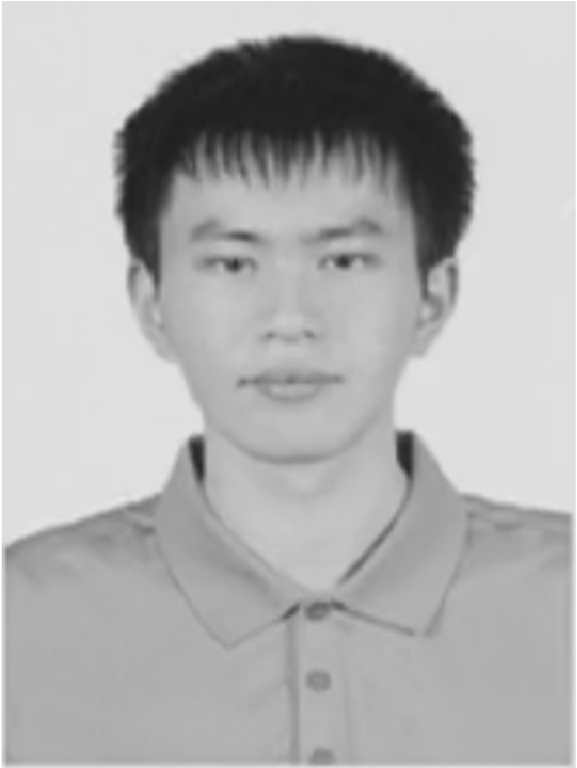}}]
{Xinyu Wang} received his BS degree in Computer Science from Beijing Institute of Technology in 2022.
He is currently a Ph.D. student at the School of Computer Science and Technology at the Beijing Institute of Technology, China.
His major research interests include information visualization.
\end{IEEEbiography}

\begin{IEEEbiography}
[{\includegraphics[width=1in,height=1.25in,clip,keepaspectratio]{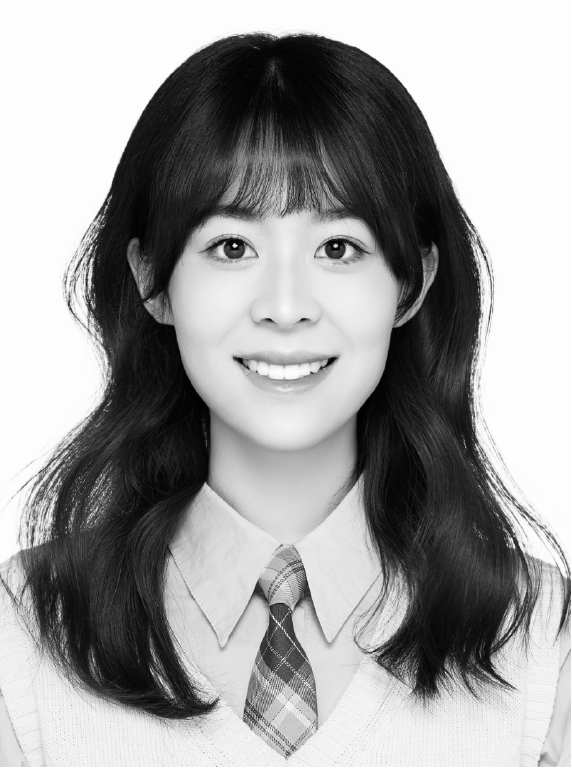}}]
{Runfei Li} received her BS degree in Computer Science from Beijing Institute of Technology in 2022. 
She is currently a master student at the School of Computer Science and Technology at the Beijing Institute of Technology, China.
Her major research interests include data visualization and human-computer interaction.
\end{IEEEbiography}

\begin{IEEEbiography}
[{\includegraphics[width=1in,height=1.25in,clip,keepaspectratio]{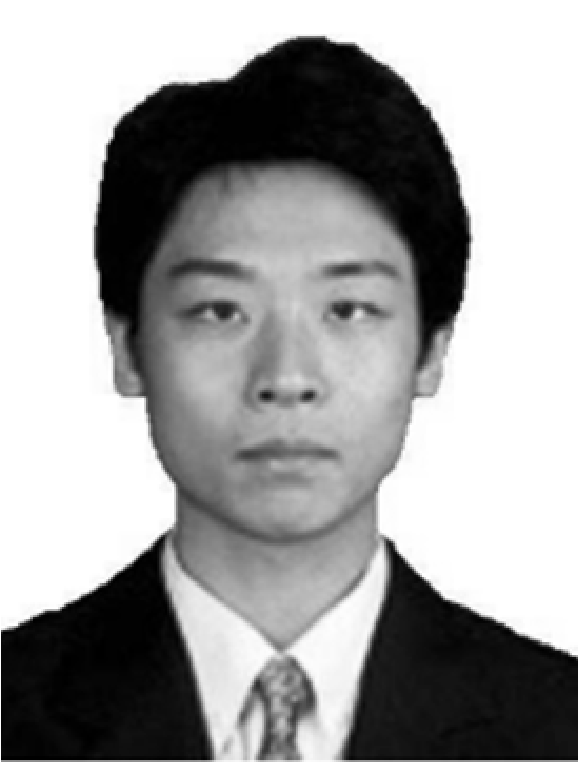}}]
{Chi Harold Liu} received his B.Eng. degree from Tsinghua University, Beijing, China, in 2006, and the Ph.D. degree from the Imperial College London, London, U.K., in 2010. He is currently a Full Professor and the Vice Dean with the School of Computer Science and Technology, Beijing Institute of Technology, China. Before moving to Academia, he joined IBM Research, China, Beijing, as a Staff Researcher and a Project Manager, after working as a Post-Doctoral Researcher with the Deutsche Telekom Laboratories, Berlin, Germany, and a Visiting Scholar with the IBM T. J. Watson Research Center, Yorktown Heights, NY, USA. He has published more than 90 prestigious conference and journal articles and holds more than 14 EU/U.S./U.K./China patents. His current research interests include the big data analytics, mobile computing, and deep learning. Dr. Liu is a fellow of IET.
\end{IEEEbiography}

\begin{IEEEbiography}
[{\includegraphics[width=1in,height=1.25in,clip,keepaspectratio]{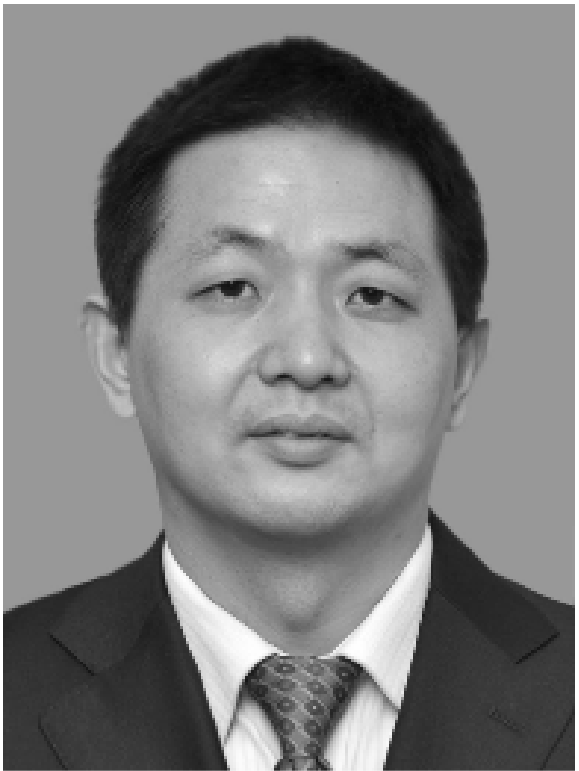}}]
{Chuangxin Ou} is a data scientist with PICC Information Technology Company Limited., China. His research interests include financial intelligence, data cleaning and visualization.
\end{IEEEbiography}

\begin{IEEEbiography}
[{\includegraphics[width=1in,height=1.25in,clip,keepaspectratio]{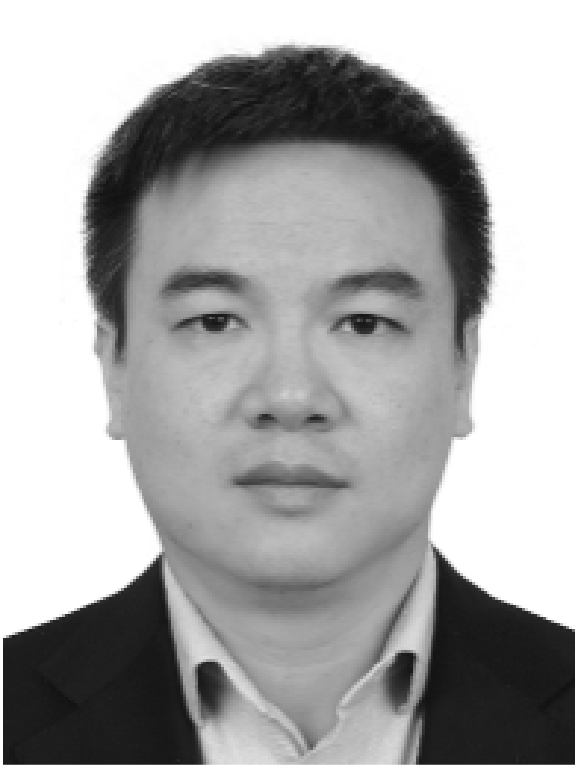}}]
{Dong He} is a data scientist with PICC Information Technology Company Limited., China. His research interests include data management, data mining and visualization.
\end{IEEEbiography}

\begin{IEEEbiography}
[{\includegraphics[width=1in,height=1.25in,clip,keepaspectratio]{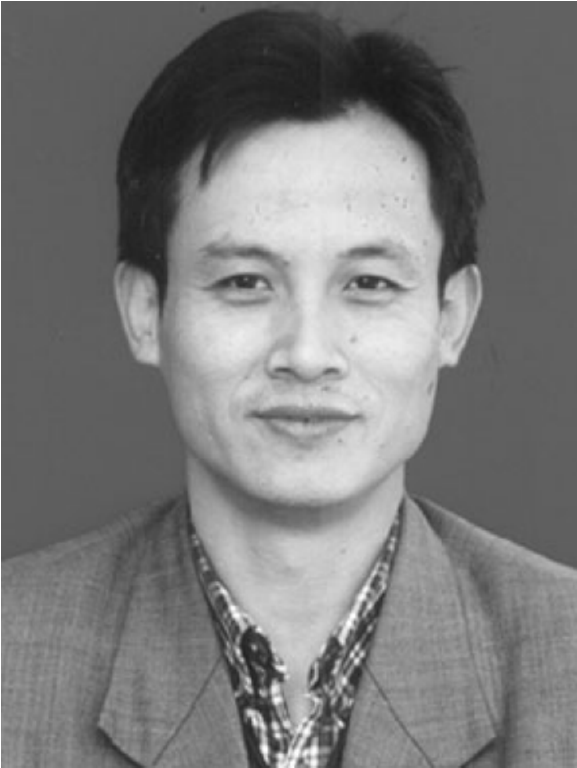}}]
{Guoren Wang} received the BSc, MSc, and PhD degrees in computer science from Northeastern University, China, in 1988, 1991, and 1996, respectively. Currently, he is a professor with the Beijing Institute of Technology, China. His research interests include XML data management, query processing and optimization, bioinformatics, high-dimensional indexing, parallel database systems, and P2P data management.
\end{IEEEbiography}

\end{document}